\documentstyle[apjgalley,psfig,multicol]{article}
\font\fiverm=cmr5             \font\sevenrm=cmr7

          \font\sixrm=cmr6       
                     
\def\teq#1{$\, #1\,$}                           
\def\erg{\varepsilon}

\def\lambar{\lambda\llap {--}}
\def\fsc{\alpha_{\hbox{\sevenrm f}}}                                
\def\dover#1#2{\hbox{${{\displaystyle#1 \vphantom{(} }\over{
   \displaystyle #2 \vphantom{(} }}$}}

\def\ecyc{\omega_{\hbox{\fiverm B}}}
\def\today{\ifcase\month\or
  January\or February\or March\or April\or May\or June\or
  July\or August\or September\or October\or November\or
  December\fi
  \space\number\day, \number\year}
\begin{document}
%
%
\newcommand{\vol}[2]{$\,$\rm #1\rm , #2.}                 
\newcommand{\figureout}[5]{\centerline{}
   \centerline{\hskip #3in \psfig{figure=#1,width=#2in}}
   \vspace{#4in} \figcaption{#5} }
\newcommand{\twofigureout}[3]{\centerline{}
   \centerline{\psfig{figure=#1,width=3.4in}
        \hskip 0.5truein \psfig{figure=#2,width=3.4in}}
    \figcaption{#3} }    

\newcommand{\figureoutfixed}[2]{\centerline{}
   \centerline{\psfig{figure=#1,width=5.5in}}
    \figcaption{#2}\clearpage }
\newcommand{\figureoutsmall}[2]{\centerline{\psfig{figure=#1,width=5.0in}}
    \figcaption{#2}\clearpage }
\newcommand{\figureoutvsmall}[2]{\centerline{\psfig{figure=#1,width=4.0in}}
    \figcaption{#2}\clearpage }
\newcommand{\tableout}[4]{\vskip 0.3truecm \centerline{\rm TABLE #1\rm}
   \vskip 0.2truecm\centerline{\rm #2\rm}   
   \vskip -0.3truecm  \begin{displaymath} #3 \end{displaymath} 
   \noindent \rm #4\rm\vskip 0.1truecm } 
%
%
%
%

\title{PHOTON SPLITTING AND PAIR CREATION 
       IN HIGHLY MAGNETIZED PULSARS}

   \author{Matthew G. Baring\altaffilmark{1} and Alice K. Harding}
   \affil{Laboratory for High Energy Astrophysics, Code 661, \\
      NASA Goddard Space Flight Center, Greenbelt, MD 20771, U.S.A.\\
      \it baring@lheavx.gsfc.nasa.gov, harding@twinkie.gsfc.nasa.gov\rm}
   \altaffiltext{1}{Universities Space Research Association}
   \authoraddr{Laboratory for High Energy Astrophysics, Code 661,
      NASA Goddard Space Flight Center, Greenbelt, MD 20771, U.S.A.}
\slugcomment{To appear in \it The Astrophysical Journal\rm , 
Vol 547, February 1, 2001 issue.}
%

\begin{abstract}  
The absence of radio pulsars with long periods has lead to the popular
notion of a high \teq{P} ``death line.''  In the standard picture,
beyond this boundary, pulsars with low spin rates cannot accelerate
particles above the stellar surface to high enough energies to initiate
pair cascades, and the pair creation needed for radio emission is
strongly suppressed.  In this paper we explore the possibility of
another pulsar ``death line'' in the context of polar cap models,
corresponding to high magnetic fields B in the upper portion of the
period-period derivative diagram, a domain where few radio pulsars are
observed.  The origin of this high B boundary, which may occur when B
becomes comparable to or exceeds $B_{\rm cr} = 4.4 \times 10^{13}$
Gauss, is also due to the suppression of magnetic pair creation, but
primarily because of ineffective competition with magnetic photon
splitting.  Threshold pair creation also plays a prominent role in the
suppression of cascades.  We present Monte Carlo calculations of the
pair yields in photon splitting/pair cascades which show that, in the
absence of scattering effects, pair production is effectively
suppressed, but only if all three modes of photon splitting allowed by
QED are operating in high fields.  This paper describes the probable
shape and position of the new ``death line,'' above which pulsars are
expected to be radio quiet, but perhaps still X-ray and gamma-ray
bright.  The hypothesized existence of radio-quiet sources finds
dramatic support in the recent discovery of ultra-strong fields in Soft
Gamma-ray Repeaters and Anomalous X-ray Pulsars.  Guidelines for
moderate to high B pulsar searches at radio wavelengths and also in the
soft and hard gamma-ray bands are presented.
\end{abstract}  
\section{INTRODUCTION}
\label{sec:intro}

Most current theories for the generation of coherent radio emission in
pulsar magnetospheres (see summations in Michel 1991; Melrose 1993;
Lyutikov et al. 1999) require formation of an electron-positron pair
plasma.  Such a pair plasma has been shown to develop via
electromagnetic cascades along the open magnetic field lines as a
result of particle acceleration to TeV energies, followed by curvature,
synchrotron radiation and one-photon pair production in the strong
magnetic field, \teq{\gamma\to e^+e^-}, near the neutron star surface
(e.g. Sturrock 1971; Ruderman \& Sutherland 1975, Arons \& Scharlemann
1979).  Numerical simulations of pair cascades above a pulsar polar cap
show that \teq{10^3-10^4} pairs are produced for each accelerated
primary electron (Daugherty \& Harding 1982) in a typical young
pulsar.  But if any of the necessary conditions for development of a
pair cascade, {\it i.e.} production of $\gamma$-ray photons (requiring
particle acceleration to high energy) or production of sufficient
pairs, is absent the pulsar should not, according to theory, emit
detectable radio emission.  The existence of an observed ``death line''
in the \teq{P-\dot P} distribution of known radio pulsars (see
Figure~\ref{fig:quiescence} below), to the right of which no pulsars
were detected, until the recent discovery (Young, Manchester \&
Johnston 1999) of the 8.5 second pulsar PSR J2144-3933, provided some
circumstantial evidence that pairs are required for radio emission. The
slope of this putative boundary fits a line of constant voltage across
the open field lines, \teq{V \propto P^{-3/2} \dot P^{1/2}}, suggesting
that radio emission ceases when the particle acceleration is not high
enough to sustain pair production.  The radio emission from some long
period pulsars like PSR J2144-3933 can be explained through production
of pairs by photons from inverse-Compton scattering (Arons 2000, Zhang,
Harding \& Muslimov 2000), a process which does not require as high a
voltage as does curvature radiation.

In this paper, we study another area of pulsar phase space, that of
very high magnetic fields, where the character of pair cascades is
significantly altered by the process of photon splitting and other effects
such as ground state pair creation and positronium formation.  Magnetic
photon splitting, \teq{\gamma\to\gamma\gamma}, a QED process in which a
single photon splits into two lower-energy photons (Adler 1971, Baring
\& Harding 1997), operates efficiently and competes effectively with
pulsar pair production only in magnetic fields above \teq{\sim
10^{13}}Gauss (Harding, Baring \& Gonthier 1997, hereafter HBG97).
This region of high magnetic field strength lies in the upper-right
part of the \teq{P-\dot P} diagram.  We will explore whether photon
splitting may be the explanation for why there are no classical radio
pulsars detected with derived magnetic fields above \teq{\sim
10^{14}}Gauss. Our initial results (Baring \& Harding 1998) have
suggested that photon splitting could indeed explain the absence of
very high-field radio pulsars.  This paper considers in more detail the
underlying assumptions made in that calculation, explores the physics
of high-field pair cascades and makes predictions for high-energy
searches for high-field pulsars.  Although pulsars in this region of
phase space may be radio-quiet, those that are to the left of the death
line are still prolifically accelerating particles to high energies.
Their pulsed emission may therefore be detectable in the X-ray and
gamma-ray wavebands, making this relatively-unexplored region of pulsar
phase space extremely interesting.

There has recently been growing evidence for another class of pulsars
with ultra-strong magnetic fields.  Observations at X-ray energies have
yielded detections of both long periods and high period derivatives in
two types of sources, anomalous X-ray pulsars (AXPs) and soft
$\gamma$-ray repeaters (SGRs), which suggest dipole spin-down fields in
the range \teq{10^{14} - 10^{15}} Gauss.  The AXPs are a group of seven
or eight pulsating X-ray sources with periods in the range 6--12
seconds, and are continuously spinning down (Vasisht \& Gotthelf
1997).  The SGRs are a type of $\gamma$-ray transient source that
undergoes repeated bursts; four (possibly now five: see Cline et al.
2000 for SGR 1801-23) are currently known to exist.  A detailed
discussion of these sources and recent discoveries pertaining to them
is presented in Section~\ref{sec:magnetars}.  While conventional radio
pulsars tap their rotational energy to power their emission, if such a
new class of ultra-magnetized neutron stars or ``{\it magnetars}'' does
exist, their long periods indicate that their emission cannot be
rotationally-powered: they may instead be driven by their magnetic
energy.   With the possible exception of the recent report (Shitov
1999; Shitov, Pugachev \& Kutuzov 2000) of a low frequency pulsed
detection of a counterpart (PSR J1907+0919 at 111 MHz) to the soft
gamma repeater SGR 1900+14, which has not been confirmed by pulsation
searches in Aricebo data at higher frequencies (Lorimer \& Xilouris
2000) none of these sources has detectable pulsed radio emission.

In several previous papers, we have computed the photon splitting
attenuation lengths of photons emitted parallel, or at small angles, to
the magnetic field near the surface of strongly magnetized neutron
stars and compared them to the attenuation lengths for one-photon pair
production (Harding, Baring \& Gonthier 1996; 1997).  Both processes
depend on the photon energy \teq{\omega} and the angle \teq{\theta_{\rm
kB}} of photon propagation to the local field.  However, since photon
splitting has no threshold, it may occur before the photon reaches the
threshold for pair production at \teq{\omega \sin\theta_{\rm kB} =
2m_ec^2}.  We found that the photon splitting attenuation lengths are
lower than those for pair production in fields higher than \teq{\sim 3
\times 10^{13}}Gauss, for emission in the open field region near the
neutron star surface.  One must therefore incorporate photon splitting
in cascade models of highly-magnetized pulsars.  This has been done in
the case of PSR1509-58 (Harding, Baring \& Gonthier 1997), which has a
field of \teq{3\times 10^{13}}Gauss as estimated using
Equation~(\ref{eq:spindown}).  We considered both the case where all
{\it linear} polarization modes of photon splitting allowed by CP
invariance operate, and the case where only the \teq{\perp \rightarrow
\parallel\parallel} mode (our polarization convention is defined near
the beginning of Section~\ref{sec:suppression}) allowed by selection
rules (Adler 1971) in the weakly dispersive limit operates.  The latter
case, we argued, was most appropriate for pulsars.  In both cases, we
found that photon splitting attenuation is a plausible explanation for
the very low spectral cutoff between 10 and 30 MeV in PSR1509-58
(Kuiper et al. 2000).  Since the rates of these attenuation mechanisms
are very sensitive to the polarization states of the incoming photons,
it is evident that the polarization properties of the radiated photons
are very important in modeling both the output spectrum and pair
suppression in high-field pulsars.

In ultra-strong magnetic fields, the vacuum dispersion increases to the
point where the selection rules derived for weak linear dispersion 
(appropriate for \teq{B\lesssim B_{\rm cr}}) may
no longer be valid, and higher order non-linear contributions to such
dispersion may be significant.  We presently do not know which modes of
photon splitting operate in supercritical fields and thus do
not know which photon polarization modes are allowed to split.  We will
therefore explore several possibilities for photon splitting mode
behavior in our study of high-magnetic field pair cascades.  In
Section~\ref{sec:suppsplit}, we discuss the physics of pair suppression
in high magnetic fields, both by photon splitting and by pair
production in low-lying Landau states, as well as the possible
alternative mechanism of pair suppression by bound-state pair creation,
discussed by Usov \& Melrose (1996).  In Section~\ref{sec:noradio}, we
present results of detailed simulations of splitting/pair cascades
which explore the conditions for suppression of pair creation in a
pulsar magnetosphere.  We compute the boundary in the pulsar
\teq{P-\dot P} distribution where the escape energies for photon
splitting and magnetic pair production for photons of \teq{\perp}
polarization are equal.  This condition then defines a putative
radio-quiescence boundary, posited in Baring \& Harding (1998), above
which pulsars may not produce the dense pair plasmas required for radio
emission.  For reasonable assumptions about the location of particle
acceleration and the angles of the emitted photons in high-field
pulsars, the computed radio-quiescence boundary lies at \teq{\dot P}
above those in the known radio pulsar population, potentially
explaining the absence of radio pulsars with fields above \teq{\sim
10^{14}}Gauss.  Our cascade simulations indicate that such a boundary
is viable only if photons of both polarizations can split; otherwise
pair creation is generally postponed a generation rather than
suppressed.  In the latter case, significant reductions in pair
production can ensue only if the maximum energy of primary photons is
below or not much above the pair creation escape energy.  If pair
suppression is rife at these high \teq{\dot P}, it suggests the
possible existence of a new class of radio-quiet, ultra-magnetized
pulsars that may be related to the emerging class of magnetars.

\section{PHYSICAL EFFECTS INFLUENCING PAIR CREATION}
\label{sec:suppression}

The most important competitor to pair creation \teq{\gamma\to e^+e^-}
at high magnetic field strengths as a mechanism for attenuating photons
in pulsar magnetospheres is magnetic photon splitting
\teq{\gamma\to\gamma\gamma}.  Hence this process motivated the
suggestion (Baring \& Harding 1998) that splitting can potentially
suppress pair creation and inhibit radio emission in highly-magnetized
pulsars; accordingly it forms the centerpiece of the discussion of this
Section.  Yet, the creation of pairs in low Landau levels or the
ground state can also suppress subsequent pair creation, so this effect
is addressed in Section~\ref{sec:suppground}.  We also include a brief
discussion of the role of positronium formation, since the residence of
pairs in neutral bound states can inhibit a range of coherent
mechanisms for producing radio emission.  The section concludes with a
discussion of how the combined effects of photon splitting and
threshold pair creation change the nature of pair cascades as the
magnetic field increases.

At this point, it is appropriate to identify conventions adopted
throughout this paper.  The photon linear polarizations are such that
\teq{\parallel} refers to the state with the photon's {\it electric}
field vector parallel to the plane containing the magnetic field and
the photon's momentum vector, while \teq{\perp} denotes the photon's
electric field vector being normal to this plane.  Furthermore, all
photon and electron energies will be written in dimensionless form,
being scaled by \teq{m_ec^2}.  Since general relativistic effects will
play an important role in our considerations, we adopt the convention
of labelling photon energies in the local inertial frame of reference
by \teq{\omega}, when a connection to quantum processes is most
salient, and photon energies that an observer at infinity would measure
will be denoted by \teq{\erg}.  By the
same token, \teq{B} will be used to denote local fields, and \teq{B_0}
will represent the surface field that would be inferred at the pole
{\it in flat spacetime}, i.e. a neutron star of radius \teq{R} with a
pure dipole field has a magnetic moment of \teq{\mu_{\fiverm B} = B_0R^3/2}.

Photons of \teq{\perp} polarization dominate the photon population
generated by the principal primary and secondary emission processes,
namely cyclotron/synchrotron radiation, curvature emission and resonant
Compton scattering.  Synchrotron and curvature radiation are central to
the hard X-ray and gamma-ray emission of polar cap models (e.g.
Daugherty and Harding 1982; 1996), and are essentially identical in
their polarization properties in the classical picture.  For
monoenergetic electrons, polarization levels \teq{{\cal
P}=\vert\dot{n}_{\perp}- \dot{n}_{\parallel}\vert
/\vert\dot{n}_{\perp}+\dot{n}_{\parallel}\vert} of between \teq{50\%}
and \teq{100\%} are achieved (e.g. see Fig.~6.7 of Bekefi 1966), while
for power-law electrons, the degree of polarization is generally in the
60\%--70\% range (e.g. Bekefi 1966; Rybicki and Lightman 1979), with
\teq{\perp} photons dominating in both cases.  At \teq{B\gtrsim B_{\rm
cr}}, there is some degree of quantum depolarization, though in reality
this is generally small since photon angles with respect to local
fields are always small.  The dominance of \teq{\perp} photons applies
also to the products of resonant Compton scattering, which has more
recently been considered (e.g. Sturner and Dermer 1994) as a primary
emission mechanism.  In the Thomson limit, the cross-sections for
scatterings of polarized photons can be derived from the results of
Herold (1979), indicating that \teq{\perp} photons are produced at 3
times the rate of \teq{\parallel} photons.

\subsection{Competition with Magnetic Photon Splitting}
\label{sec:suppsplit}

\subsubsection{Photon Splitting}
\label{sec:psplit}

The relevance of photon splitting
\teq{\gamma\to\gamma\gamma} to neutron star environments was emphasized
by Adler (1971), Mitrofanov et al. (1986) and Baring (1988).  In the
context of pulsars, it has been discussed by Baring (1993), who
proposed it as a mechanism for suppressing the appearance of
\teq{e^+e^-} annihilation lines, and by Harding, Baring and Gonthier
(1996, 1997) and Chang, Chen and Ho (1996), who focused on its action
in attenuating gamma-ray pulsar continua.  Splitting is a third-order
QED process with a triangular Feynman diagram.  Though it is permitted
by energy and momentum conservation, when \teq{B=0} it is forbidden by
a charge conjugation symmetry of QED known as Furry's theorem (e.g.
Jauch and Rohrlich 1980), which states that ring diagrams that have an
odd number of vertices with only external photon lines generate
interaction matrix elements that are identically zero.  This symmetry,
which pertains to the electron/positron propagators, is broken by the
presence of an external field.  The splitting of photons is therefore a
purely quantum effect, and has appreciable reaction rates only when the
magnetic field is at least a significant fraction of the quantum
critical field \teq{B_{\rm cr}}.

It is practical to
restrict considerations of splitting to regimes of weak dispersion,
where manageable expressions for its rates are obtainable, but are
still complicated triple integrations (e.g. see Adler 1971; Stoneham
1979; Baier, Mil'shtein, \& Shaisultanov 1996; Adler \& Schubert 1996)
or triple summations (Baring 2000).  Further specialization to either
low magnetic fields (\teq{B\ll B_{\rm cr}}) or low photon energies
(\teq{\omega\ll 2}) therefore proves expedient, and palatable results
for splitting rates were first obtained in such regimes by
Bialynicka-Birula and  Bialynicki-Birula (1970), Adler \it et al. \rm
(1970) and Adler (1971).  A compact presentation of these rates (i.e.
for \teq{\omega\ll 2}) for the three polarization modes of splitting
permitted by CP (charge-parity) invariance in QED, namely
\teq{\perp\to\parallel\parallel}, \teq{\perp\to\perp\perp} and
\teq{\parallel\to\perp\parallel}, is (e.g. see HBG97, who also display
differential rates)
\begin{eqnarray}
   T^{\rm sp}_{\perp\to\parallel\parallel}(\omega ) &=& 
   \dover{\fsc^3}{60\pi^2}\, \dover{1}{\lambar}\, 
   \biggl(\dover{B}{B_{\rm cr}}\biggr)^6\,\omega^5\; {\cal M}_1^2 \; =\; 
   \dover{1}{2}\, T^{\rm sp}_{\parallel\to\perp\parallel}
  \nonumber\\[-5.5pt]
  &&  \label{eq:splittatten} \\[-5.5pt]
   T^{\rm sp}_{\perp\to\perp\perp} &=& 
   \dover{\fsc^3}{60\pi^2}\, \dover{1}{\lambar}\, 
   \biggl(\dover{B}{B_{\rm cr}}\biggr)^6\,\omega^5\; {\cal M}_2^2\nonumber
\end{eqnarray}
in the frame where photons propagate perpendicular to the field, where 
\begin{eqnarray}
   {\cal M}_1 &=& \biggl(\dover{B}{B_{\rm cr}}\biggr)^{-4}
   \int^{\infty}_{0} \dover{ds}{s}\, e^{-s\, B_{\rm cr}/B}\,\nonumber\\
    &&\times\;\;
   \Biggl\{ \biggl(-\dover{3}{4s}+\dover{s}{6}\biggr)\,\dover{\cosh s}{\sinh s} 
   +\dover{3+2s^2}{12\sinh^2s}+\dover{s\cosh s}{2\sinh^3s}\Biggr\}\;\; ,
  \nonumber\\[-5.5pt]
  &&   \label{eq:splitcoeff} \\[-5.5pt]
   {\cal M}_2 &=& \biggl(\dover{B}{B_{\rm cr}}\biggr)^{-4}
   \int^{\infty}_{0} \dover{ds}{s}\, e^{-s\, B_{\rm cr}/B}\,\nonumber\\
    &&\times\;\;
   \Biggl\{ \dover{3}{4s}\,\dover{\cosh s}{\sinh s} +
   \dover{3-4s^2}{4\sinh^2s} - \dover{3s^2}{2\sinh^4s}\Biggr\}\nonumber
\end{eqnarray}
and \teq{\fsc} is the fine structure constant and \teq{\lambar
=\hbar/(m_ec)} is the Compton wavelength of the electron divided by
\teq{2\pi}.  At low fields, ${\cal M}_1$ and ${\cal M}_2$ are
independent of $B$, but at high fields possess ${\cal M}_1\propto
B^{-3}$ ${\cal M}_2\propto B^{-4}$ dependences.  These are the
expressions used in this paper because of their broad applicability to
the pulsar problem.  Deviations from this low energy limit near pair
creation threshold are presented in detail by Baier, Mil'shtein, \&
Shaisultanov (1996) and Baring \& Harding (1997), and are mentioned
below where appropriate.  An analytic approximation to the total rate
in the \teq{B\gg B_{\rm cr}} limit is presented in Baring and Harding
(1997), and \teq{B\gg B_{\rm cr}} analytic forms for other splitting
modes can be found in Baring (2000).

The birefringence of the magnetized vacuum implies an alteration of the
kinematics of strong field QED processes (Adler 1971), admitting the
possibility of non-collinear photon splitting.  Hence, while the
splitting modes \teq{\perp\to\perp\parallel},
\teq{\parallel\to\perp\perp} and \teq{\parallel\to\parallel\parallel}
are forbidden by CP invariance in the limit of zero dispersion,
dispersive effects guarantee a small but non-zero probability for the
\teq{\perp\to\perp\parallel} channel.  Extensive discussions of linear
dispersion in a magnetized vacuum are presented by Adler (1971) and
Shabad (1975); considerations of plasma dispersion (e.g. see the
analysis of Bulik 1998) are not relevant to classical gamma-ray pulsars
because of the relatively low densities present in the magnetosphere,
but may be quite pertinent to soft gamma repeaters.  Adler (1971)
showed that in the limit of {\it weak linear vacuum} dispersion
(roughly delineated by $B\sin\theta_{\rm kB} \lesssim B_{\rm cr}$),
where the refractive indices for the polarization states are very close
to unity, energy and momentum could be simultaneously conserved only
for the splitting mode \teq{\perp\to\parallel\parallel} (of the modes
permitted by CP invariance) below pair production threshold.  This
kinematic selection rule was demonstrated for linear dispersion, a
regime that applies to subcritical fields.  Therefore, it is probable
that only the one mode (\teq{\perp\to\parallel\parallel}) of splitting
operates in normal pulsars.  However, this constraint may not hold
in supercritical fields where strong vacuum dispersion arises, thereby
requiring a revised assessment using the generalized vacuum
polarizability tensor (i.e. including quadratic and higher order
contributions to the vacuum polarization).

\clearpage

\subsubsection{Attenuation Lengths}
\label{sec:attenl}

As pair creation is a first order QED process, whereas splitting is
third-order in the fine structure constant \teq{\fsc =e^2/(\hbar c)},
it is not immediately obvious that \teq{\gamma\to\gamma\gamma} can ever
dominate \teq{\gamma\to e^+e^-} in pulsar environs.  Yet it is the
propagation of the photons at small angles to the field through the
pulsar magnetosphere that affords photon splitting an opportunity to
compete effectively with pair creation when \teq{B\gtrsim B_{\rm cr}},
since the pair threshold \teq{\omega\sin\theta_{\rm kB}=2} is always
crossed from below.  An approximate assessment of the relative
importance of magnetic photon splitting and pair creation
\teq{\gamma\to e^+e^-} was performed by Harding, Baring and Gonthier
(1997) by computing {\it attenuation lengths} \teq{L} for each of these
processes.  These are the scalelengths for attenuation in the neutron
star magnetosphere, and are defined to be path lengths over which the
optical depth
\begin{equation} 
   \tau(\Theta, \erg,\, l) = \int_0^l T(\theta_{\rm kB}, \omega ) ds
\label{eq:tau}
\end{equation}
is unity, i.e. \teq{\tau(\Theta, \erg,\, L)=1}.  In computing such
attenuation lengths, it is essential to fully include the effects of
general relativity.  The reason for this is that the quantum transition
rates for both splitting and pair creation are strong functions of the
photon energy and angle and the magnitude of the magnetic field
strength in the local inertial frame, all of which are strongly
influenced by curved spacetime.  Such an analysis is confined to the
Schwarzschild metric because the dynamical timescales for gamma-ray
pulsars are considerably shorter than their period (e.g.
\teq{P=0.15}sec. for PSR1509-58), so that rotation effects in the Kerr
metric, for example those due to frame-dragging, can be neglected in
our photon attenuation analysis.  Throughout this paper, we assume a
neutron star mass, \teq{M = 1.4\,M_\odot} and radius, \teq{R=10^6}cm.

Values for \teq{L} are computed assuming that test photons are emitted
on or above the neutron star surface at some polar colatitude
\teq{\Theta} (usually chosen to be at the polar cap rim) and propagate
outward, initially at a specified angle \teq{\theta_{\rm kB,0}} to the
gravitationally-modified dipole magnetic field; a depiction of the
geometry is presented in HBG97.  Frequently in this paper, a surface
origin of the photons is chosen to provide a concise and representative
presentation of the attenuation properties of the pulsar
magnetosphere.  Such attenuation lengths possess power-law behavior at
high energies, with \teq{L\propto \erg^{-5/7}} for photon splitting,
and \teq{L\simeq 2/\erg} for pair creation just above threshold (HBG97;
proportionalities that hold in both curved and flat spacetime).
Furthermore, they display the property that the attenuation length
declines with colatitude of emission, a consequence of the associated
increase in curvature of the field lines.  At low energies, the
attenuation lengths diverge and photons escape the magnetosphere
without attenuation.  The \teq{L\to\infty} asymptotes define {\it
escape energies} (HBG97), below (above) which spectral transparency
(opacity) is effectively guaranteed.  The existence of such escape
energies is a consequence of the \teq{r^{-3}} decay of the dipole
field: there are always sufficiently low photon energies for which the
field declines before the photons have had sufficient time to attenuate
in their passage through the magnetosphere.

\subsubsection{Escape Energies}
 \label{sec:erg_escape}

\begin{figure*}[t]
\twofigureout{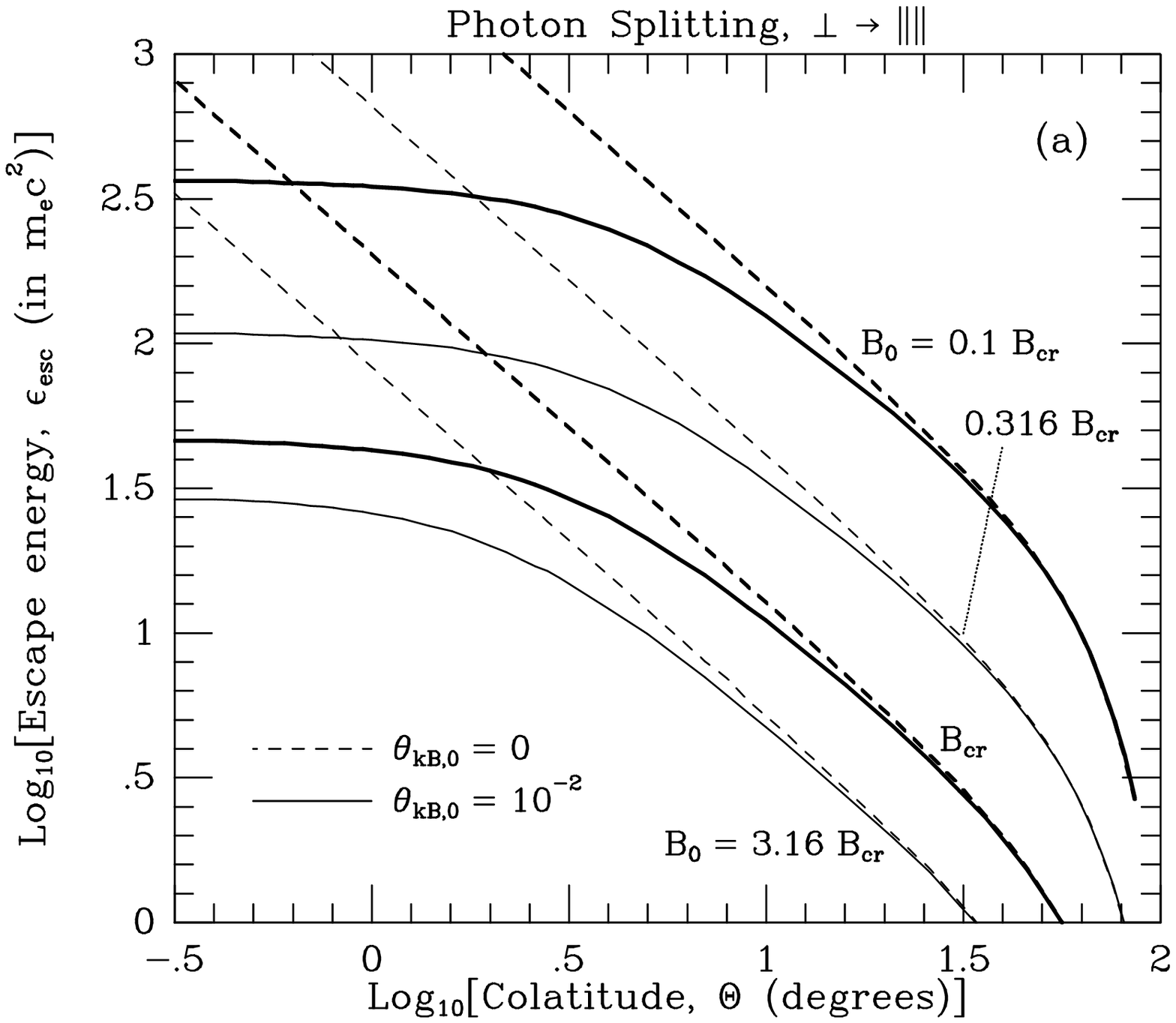}{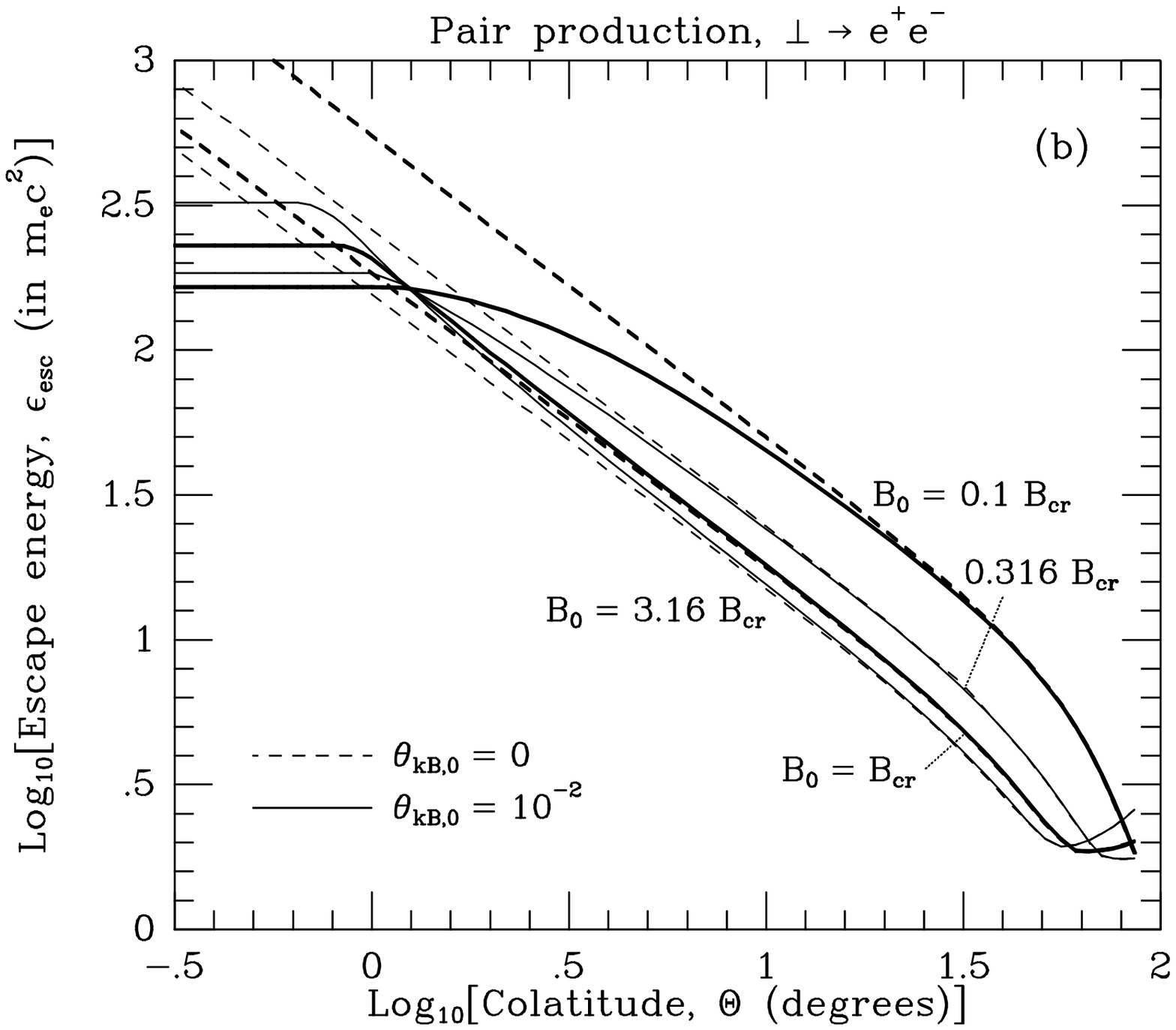}{
 The escape energy (i.e. where \teq{L\to\infty}) for (a) photon
 splitting and (b) pair production as a function of magnetic colatitude
 for photon emission both along \teq{{\bf B}} (dashed curves) and at
 angle \teq{\theta_{\rm kB,0}=0.01} radians (\teq{=0.57^\circ}) to the
 field (solid curves).  The escape energies for each process are
 monotonically decreasing functions of \teq{B} for the range of
 parameters shown.  The \teq{\theta_{\rm kB,0}=0} curves have slopes of
 -6/5 (splitting) and -1 (pair creation) at small \teq{\Theta}, as
 discussed by Harding, Baring \& Gonthier (1997), and diverge near
 \teq{\Theta = 0}, where the field line radius of curvature becomes
 infinite.  For the solid curves, at low magnetic colatitudes
 \teq{\Theta \lesssim 10\theta_{\rm kB,0}}, the field curvature is so
 low that photon attenuation is insensitive to the value of \teq{\Theta}
 and is well described by the uniform field results in
 Eqs.~(\ref{eq:splittatten}) and~(\ref{eq:splitcoeff}).  The escape
 energies are presented for surface photon emission, \teq{R_0=R}.
 \label{fig:escape} }      
\end{figure*}

For each photon trajectory through the magnetosphere, corresponding to
a specific set of values for the colatitude \teq{\Theta} and
\teq{\theta_{\rm kB,0}} of emission at the surface, and for a
particular dipole surface field \teq{B_0}, the escape energy
\teq{\erg_{\rm esc}} is uniquely defined and cleanly delineates the
energy ranges of source opacity and transparency. The escape energy
also depends on the radius of emission $R_0$ (see Figure
\ref{fig:quiescence}).  The dependence of \teq{\erg_{\rm esc}} on such
parameters is shown in Figure~\ref{fig:escape} for photons that are
initially of polarization \teq{\perp}.  Escape energy plots for
unpolarized photons are given in Figures~3 and~5 of HBG97.  The feature
of Figure~\ref{fig:escape} that is most salient for the considerations
of this paper can be obtained by comparing the plots for splitting and
pair production. For low fields, pair production escape energies are
below those for splitting, but the situation is reversed in high
fields; the escape energies are roughly equal for a narrow band of
fields around \teq{B_0\sim 0.5 B_{\rm cr}}.  For field strengths
\teq{B_0\gtrsim B_{\rm cr}}, photon splitting can attenuate photons
well below pair threshold at significant colatitudes.  Hence splitting
can be expected to dominate \teq{\gamma\to e^{\pm}} in supercritical
fields for the attenuation of photons of \teq{\perp} polarization.

The escape energies in Figure~\ref{fig:escape} generally decline with
\teq{\Theta} and are monotonically decreasing functions of \teq{B_0}
for the range of fields shown.  Consider first the dashed curves in
each panel, corresponding to initial propagation of photons along the
field.  The divergences as \teq{\Theta\to 0} are due to the divergence
of the field line radius of curvature at the poles.  The maximum angle
\teq{\theta_{\rm kB}} achieved before the field falls off and inhibits
attenuation is proportional to the colatitude \teq{\Theta}.  For photon
splitting, since the rate in equation~(\ref{eq:splittatten}), and
therefore also the inverse of the attenuation length \teq{L}, is
proportional to \teq{\omega^5\sin^6\theta_{\rm kB}}, it follows that
the escape energy scales as \teq{\erg_{\rm esc} \propto \Theta^{-6/5}}
near the poles.  At fields \teq{B_0 \gtrsim 0.3B_{\rm cr}}, there is a
diminishing dependence of \teq{B} in the attenuation coefficient (e.g.
see Adler 1971; Baring \& Harding 1997 for graphical illustrations) so
that a saturation of the photon splitting attenuation lengths and
escape energies arises in highly supercritical fields.  For pair
production, the behaviour of the rate (and therefore \teq{1/L}) is
exponential in \teq{1/(\omega B\sin\theta_{\rm kB})} (e.g. see
Daugherty \& Harding 1983), which then quickly yields a dependence
\teq{\erg_{\rm esc} \propto \Theta^{-1}} near the poles for
\teq{B_0\lesssim 0.1B_{\rm cr}}.  This behavior extends to higher
surface fields because production then occurs at threshold, which
determines \teq{\erg_{\rm esc}\sim 2/\theta_{\rm kB} \propto
\Theta^{-1}}.  The pair production escape energy curves are bounded
below by the pair threshold \teq{2/\sin\theta_{\rm kB}} at the point of
pair creation (not photon emission), and for high \teq{\Theta} approach
the pair threshold \teq{(1+\sqrt{1+2B/B_{\rm cr}}) /\sin\theta_{\rm
kB}} (with \teq{\sin\theta_{\rm kB}\sim 1} determined by geometry),
blueshifted by the factor \teq{(1-2GM/Rc^2)^{-1/2} \sim 1.3}.  The
field dependence in this saturation energy arises because the creation
of pairs by \teq{\perp}-polarized photons cannot leave both the
electron and positron in the lowest Landau level (e.g. see Daugherty \&
Harding 1983); the minimum energy configuration requires one member of
the pair to be in the first excited state.

Since pulsar cascade high energy emission from curvature radiation,
inverse Compton or synchrotron by relativistic particles with Lorentz
factor $\gamma_e$  will not beam the photons precisely along the
magnetic field, but within some angle $\sim 1/\gamma_e$ to the field,
it is important to illustrate the effect on the escape energies of a
non-zero angle of emission \teq{\theta_{\rm kB,0}} of the photons
relative to {\bf B}.  This is done via the solid curves in
Figure~\ref{fig:escape}, where \teq{\theta_{\rm kB,0}=10^{-2}} (i.e.
\teq{0.57^\circ}) is taken {\it towards} the dipole axis; qualitatively
and quantitatively similar behavior arises when \teq{\theta_{\rm
kB,0}=10^{-2}} is taken away from this axis.  Curvature
radiation-initiated cascades generally have \teq{\gamma_e\sim 10^7}
(e.g. Daugherty \& Harding 1989; see also Harding \& Muslimov 1998),
while inverse-Compton seeded pair cascades yield \teq{\gamma_e\sim
3\times 10^5}--\teq{10^6} (e.g.  Sturner 1995; see also Harding \&
Muslimov 1998), both yielding angles \teq{\theta_{\rm kB,0}} very much
smaller than the example in the figure.  The situation is similar for
synchrotron radiation from first generation secondary electrons, where
\teq{\gamma_e\sim 10^3 - 10^4}; such Lorentz factors result from pairs
created by primary photons (curvature or resonant Compton) in the 1
GeV--10 GeV range.  However, as the cascade proceeds, higher
generations of pairs achieve lower Lorentz factors, resulting in a
cumulative power-law spectrum between \teq{\gamma_e\sim 10^2} and
\teq{\gamma_e\sim 10^4} (Daugherty \& Harding 1982).  This behaviour
applies to pulsars with low to moderate magnetic fields,
\teq{B_0\lesssim 0.2 B_{\rm cr}}, a domain that is well-studied in the
literature.  For more highly-magnetized pulsars, the restriction of
pair creation to a single state, discussed at length in
Section~\ref{sec:suppground}, will inhibit the creation of second and
higher generation pairs, implying that the dominant synchrotron signal
will sample angles to the field much smaller than \teq{\theta_{\rm
kB,0} =10^{-2}}.  Hence, in summation, we expect that \teq{\theta_{\rm
kB,0} \lesssim 10^{-4}-10^{-3}} will be representative for the
considerations of this paper.  The \teq{\theta_{\rm kB,0} =10^{-2}}
choice in Figure~\ref{fig:escape} aids clarity of illustration.

The general behaviour of the \teq{\theta_{\rm kB,0}=10^{-2}} curves in
Figure~\ref{fig:escape} can be simply understood.  For the most part,
the escape energy is insensitive to the emission angle for \teq{\Theta
\gtrsim 10 \theta_{\rm kB,0}}.  For small angles, the escape energy
decreases and the curves flatten below the \teq{\theta_{\rm kB,0} = 0}
curves, converging as \teq{\Theta\to 0} to an energy that is
proportional to \teq{(B_0\sin\theta_{\rm kB,0})^{-6/5}} when
\teq{B_0\ll B_{\rm cr}}.  This convergence is a consequence of the
field being almost uniform and tilted at about angle \teq{\theta_{\rm
kB,0}} to the photon path for trajectories that originate near the
pole; this behaviour at low colatitudes was first noted, in the case of
pair creation in flat spacetime, by Chang, Chen and Ho (1996).  These
``saturation'' energies follow from the dependence of the photon
splitting rate on \teq{B} and \teq{\Theta}, and have a declining
sensitivity to \teq{B_0} when \teq{B_0\gtrsim 0.3 B_{\rm cr}}.  In
Fig.~\ref{fig:escape}b, the same effect is seen for pair creation, but
this time the ``saturation'' is at the redshifted threshold energy for
\teq{\gamma\to e^{\pm}}, which varies as \teq{(1+\sqrt{1+2B/B_{\rm
cr}}) /\sin\theta_{\rm kB,0}}.  Hence, instead of the pair creation
escape energies always being monotonically decreasing functions of
\teq{B_0}, they experience an inversion at small colatitudes and
increase with \teq{B_0}.  Such an effect was not present in the
polarization-averaged escape energies presented in HBG97 because the
\teq{\parallel} state could always access the true pair threshold, i.e.
\teq{2/\sin\theta_{\rm kB,0}}.  Note also, that while splitting escape
energies always drop when \teq{\theta_{\rm kB,0}} is increased from
zero to \teq{10^{-2}}, opposite behaviour is seen in high fields for
pair creation at moderate colatitudes, due to subtleties concerning the
sudden onset of pair creation (HBG97).

It is important to emphasize that general relativistic effects are
crucial to these calculations.  While the dependence of the escape
energy on emission colatitude and surface polar field strength is
similar in curved and flat spacetime, the introduction of the
Schwarzschild metric results in decreases of the escape energies for
both process by factors of between 2 and 4, as is illustrated in Figure~4
of HBG97.  Moreover, these decreases differ for splitting and pair
creation, making it imperative to carefully account for propagation in
curved spacetime when comparing attenuation characteristics for the two
processes.  The largest effects of the incorporation of general
relativity are due to the increase of the surface dipole field strength
by roughly a factor of 1.4 above the dipole spin-down estimate of
\teq{B_0}, and the correction for the gravitational redshift of the
photon, which increases the photon energy by roughly a factor 1.2--1.3
in the local inertial frame at the neutron star surface compared to the
energy measured by the observer in flat spacetime at infinity.  The
influence of these modification factors is amplified by the sensitivity
of the rates for \teq{\gamma\to\gamma\gamma} and \teq{\gamma\to e^+e^-}
to \teq{B} and \teq{\erg} so that 20\%-30\% deviations from flat
spacetime parameters map over to the factors of 2--4 in the escape
energies.  Near the polar cap, the curvature of the photon trajectory
in a Schwarzschild metric does not affect the escape energies, to first
order, since it is generally compensated for by relativistic
modifications to the curvature of the magnetic field (e.g. see 
Gonthier \& Harding 1994).

\begin{figure*}[t]
\twofigureout{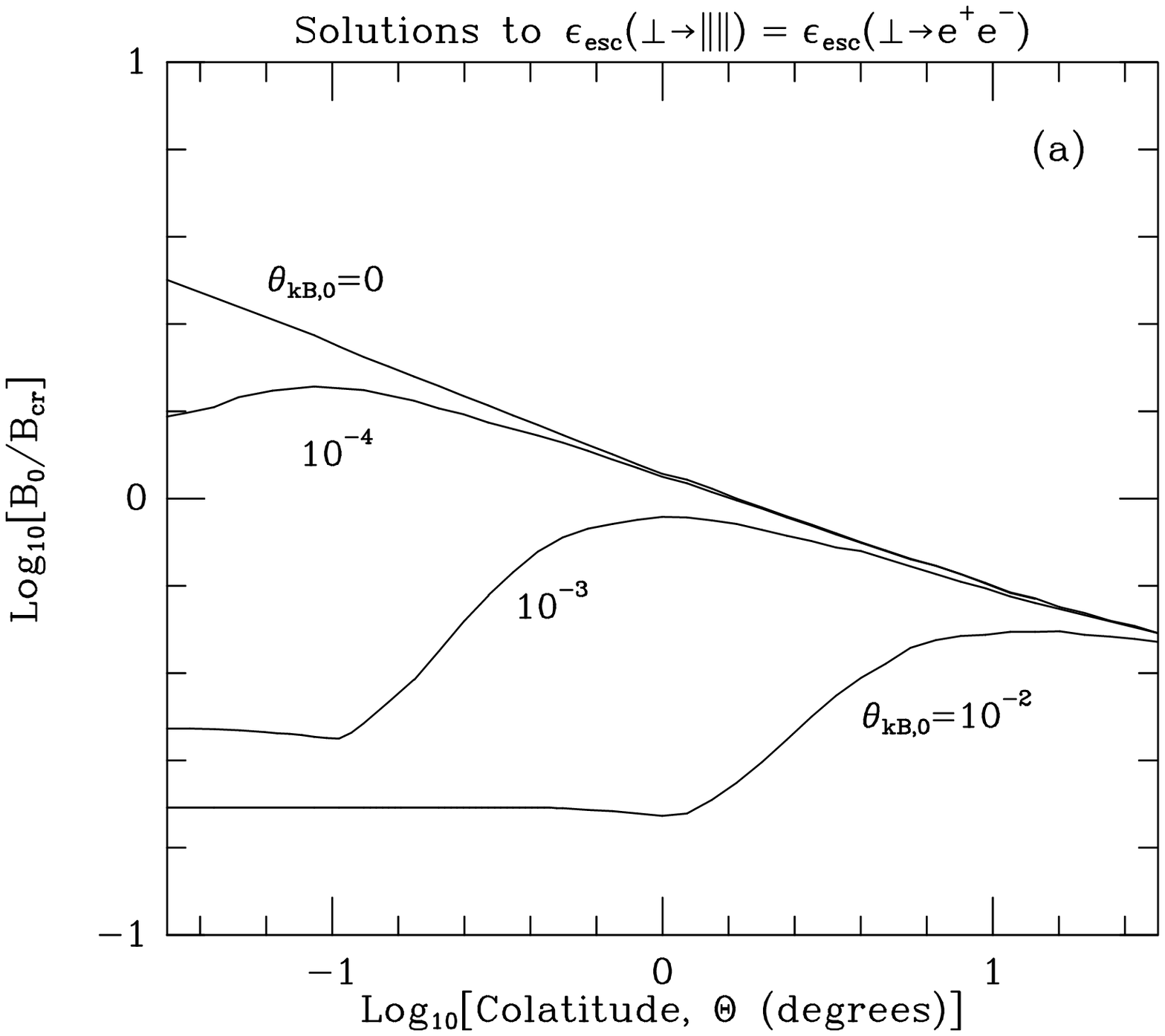}{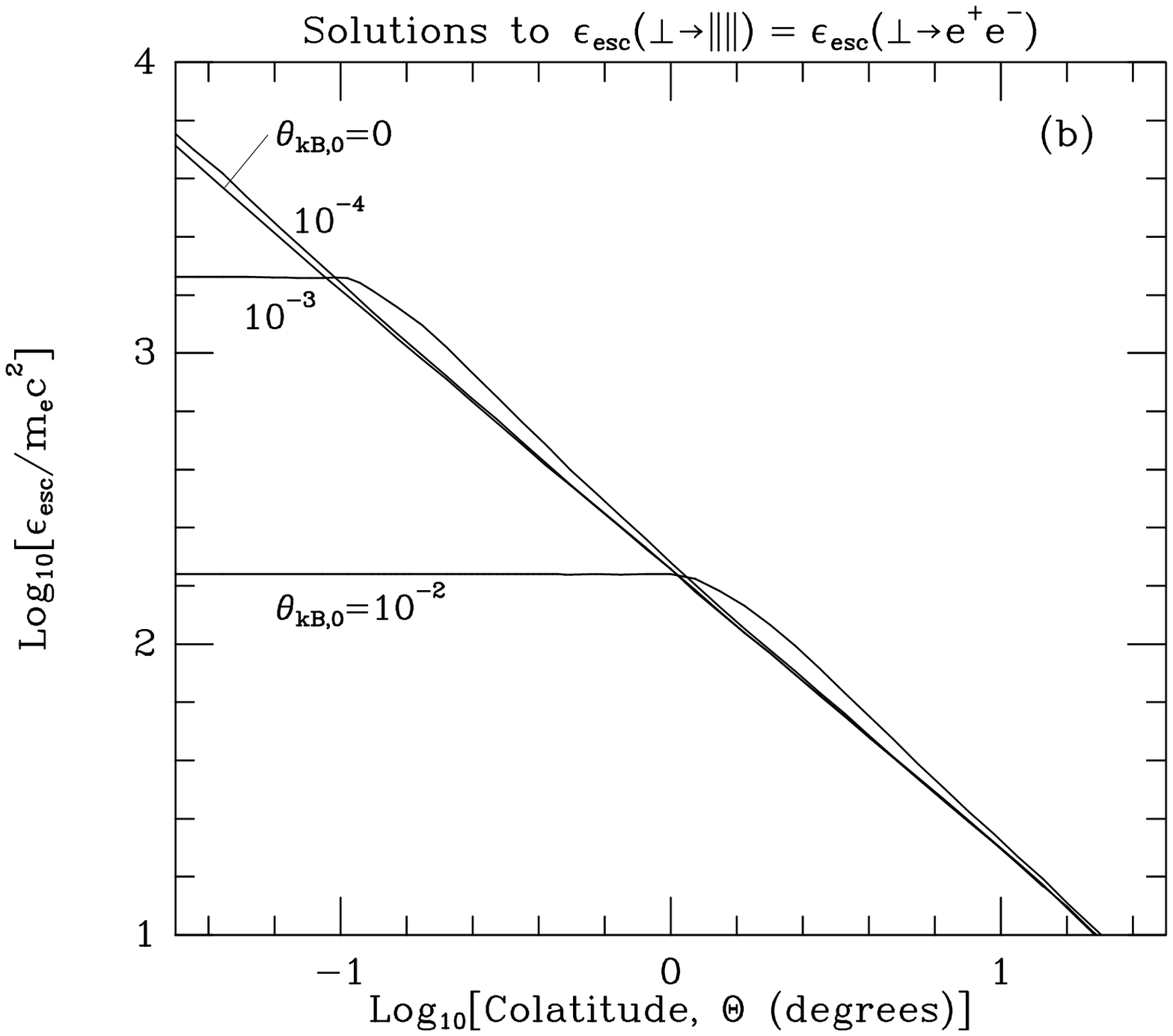}{
 The solutions to Eq.~(\ref{eq:ergescsupp}) that define (a) the polar
 surface field \teq{B_0} and (b) the escape energy \teq{\erg_{\rm esc}}
 (for either photon splittings \teq{\perp\to\parallel\parallel} or pair
 creation \teq{\perp\to e^+e^-}) as functions of the colatitude
 \teq{\Theta} of the point of photon emission, which is assumed to take
 place on the stellar surface.  Solutions are displayed for the four
 values of \teq{\theta_{\rm kB,0}}, as labelled, with the
 \teq{\theta_{\rm kB,0}=0} case generating the approximate dependences
 \teq{B_0\propto \Theta^{-4/15}} and \teq{\erg_{\rm
 esc}\propto\Theta^{-1}}.  All \teq{\theta_{\rm kB,0}\neq 0} cases
 produce \teq{B} and \teq{\erg_{\rm esc}} that are independent of
 \teq{\Theta} for sufficiently small colatitudes, namely roughly when
 \teq{\Theta\lesssim 10^4\theta_{\rm kB,0}}.
  \label{fig:ergescsol} }      
\end{figure*}

The most concise designation of when the creation of pairs
might be suppressed by photon splitting is when the escape
energies of the two processes are set equal:
\begin{equation}
   \erg_{\rm esc}^{\perp\to\parallel\parallel}(B_0,\, \Theta,\, R_0 )\; =\;
   \erg_{\rm esc}^{\perp\to e^+e^-}(B_0,\, \Theta,\, R_0 )\quad ,
 \label{eq:ergescsupp}
\end{equation}
where \teq{R_0} is the radius of the point of emission.  Note that
defining a similar equality for the \teq{\parallel} polarization mode,
or a polarization-averaged alternative, leads to only very modest
changes in the ensuing results.  Baring \& Harding (1998) contended
that Eq.~(\ref{eq:ergescsupp}) is representative of
the delineation between pulsar parameter regimes where pairs are
created in profusion and when they are present only in paucity (i.e. at
higher \teq{B_0}).  The purpose of the extensive investigation of this
paper is to determine under what conditions such a demarcation is truly
representative of a suppression of pair creation in pulsars; such an
exploration appears in Section~\ref{sec:simcas} below.  The solutions
of this equality define curves in the \teq{(B_0,\, \Theta )} plane, and
therefore specify functional dependences of the dipole spin-down
magnetic field \teq{B_0} and the escape energy (for either process) on
the colatitude \teq{\Theta} of emission.  These functions are displayed
in Figure~\ref{fig:ergescsol} for different values of the initial angle
\teq{\theta_{\rm kB,0}} of the photons relative to the field.  These
solutions define the basis for our considerations of the suppression of
pair creation in the context of radio pulsars in
Section~\ref{sec:noradio} below.

To understand the behaviour of the curves, first consider the
\teq{\theta_{\rm kB,0}=0} case.  The escape energies plotted in
Fig.~\ref{fig:escape} have the approximate dependence \teq{\erg_{\rm
esc}\propto B_0^{-\alpha}\, \Theta^{-\beta}}, with \teq{\alpha\approx
3/4} (for \teq{B_{\rm cr}\lesssim B_0\lesssim 4\, B_{\rm cr}}) and
\teq{\beta =6/5} for splitting, and \teq{\beta = 1} for pair creation.
The dependence of the pair creation escape energy on field strength for
\teq{B_0\gtrsim B_{\rm cr}} is quite weak (\teq{\alpha\approx 0}) so
that solutions to Eq.~(\ref{eq:ergescsupp}) must yield escape energies
approximately proportional to \teq{\Theta^{-1}}.  This can be
immediately folded into the photon splitting proportionality
\teq{\erg_{\rm esc}\propto B_0^{-3/4}\, \Theta^{-6/5}} to yield the
magnetic field dependence \teq{B_0\propto \Theta^{-4/15}} in
Figure~\ref{fig:ergescsol}, an approximation that is applicable for
\teq{B_{\rm cr}\lesssim B_0\lesssim 4\, B_{\rm cr}}, the range of
fields of interest to the focus of this paper.  The variation of
\teq{B_0} with \teq{\Theta} actually increases very slowly at smaller
colatitudes due to the diminishing dependence of the photon splitting
rate for the \teq{\perp\to\parallel\parallel} mode on field strength in
highly supercritical fields.  The non-zero \teq{\theta_{\rm kB,0}}
solutions for \teq{B_0} are monotonically decreasing functions of
\teq{\theta_{\rm kB,0}}, and eventually become independent of
colatitude when \teq{\Theta\lesssim 10^4\theta_{\rm kB,0}}, as is
expected from the horizontal branches of the escape energy plots in
Figure~\ref{fig:ergescsol}.  Since the escape energy solutions trace
the behaviour of the pair creation \teq{\erg_{\rm esc}}, small
increases with \teq{\theta_{\rm kB,0}} are exhibited at moderate to
high colatitudes, even though for \teq{\Theta\lesssim 10^4\theta_{\rm
kB,0}}, the solutions scale as \teq{(\theta_{\rm kB,0})^{-1}}, as
expected.

\subsection{Threshold Pair Production}
 \label{sec:suppground}

As discussed in the previous sections, pair production by photons
initially emitted at small angles to the field occurs very near
threshold in magnetic fields exceeding \teq{0.1B_{\rm cr}}.  In this
case, there are only a small number of kinematically available electron
and positron Landau states.  In very high fields, the accessible number
of excited pair states diminishes, and the pairs created by most
photons in the primary particle spectrum will occupy the ground state
(for photon polarization \teq{\parallel}) or the first excited state
(for photon polarization \teq{\perp}).  Pairs in the ground state 
cannot produce synchrotron photons.  Even pairs in the first excited
state radiate photons of energy \teq{\omega\simeq \sqrt{1+2B/B_{\rm
cr}+p_z^2}\, -1} for electron momentum \teq{p_z} parallel to the field,
which will generate very few pairs (Harding \& Daugherty 1983), and
certainly none if \teq{p_z^2 < 4(1+\sqrt{1+2B/B_{\rm cr}}\, }) for
photons of \teq{\perp} polarization.  This phenomenon thus provides
another mechanism for suppressing the number of pairs produced by
pulsar cascades.  Although pairs in the ground state could be excited
to higher Landau levels through inverse-Compton scattering of soft
X-ray photons (e.g. Zhang \& Harding 2000), we leave discussion of the
potential importance of this effect to the end of this section.  

First, we examine the effect of threshold pair production.  The
threshold photon energy in the center-of-momentum (CM) frame for
excitation into the electron-positron Landau state combination (j,k) is
\begin{equation}
   \erg_{j,k} \; =\; \sqrt{1 + 2jB/B_{\rm cr}} + \sqrt{1 + 2kB/B_{\rm cr}}
  \label{eq:erg_jk}
\end{equation}
As was discussed at the beginning of this Section, the photons
radiated in pulsar cascades are predominantly of $\perp$ polarization,
which have a pair production threshold at (0,1), i.e. one member of the
pair is in the ground state and the other is in the first excited
state.  In a uniform magnetic field \teq{B},
the number of pair states available to a photon which pair
produces at energy \teq{\omega_{\hbox{\fiverm CM}}=\omega\sin\theta_{\rm kB}}
in the CM frame (for propagation angles \teq{\theta_{\rm kB}} to the
field), is approximately (Daugherty \& Harding 1983)
\begin{eqnarray}
  N_{\rm states}(\omega_{\hbox{\fiverm CM}}, B) & \approx &\dover{1}{24}\,
   \Bigl( \dover{B_{\rm cr}}{B} \Bigr)^2\,
   \omega_{\hbox{\fiverm CM}}\, (\omega_{\hbox{\fiverm CM}} + 4)\,
   (\omega_{\hbox{\fiverm CM}} - 2)^2 \;\; ,\nonumber\\[-5.5pt]
 \label{eq:Nstates} \\[-5.5pt]
   \omega_{\hbox{\fiverm CM}} &=& \omega\sin\theta_{\rm kB}\;\; . 
 \nonumber
\end{eqnarray} 
We can assess the importance of ground-state pair creation by computing
\teq{N_{\rm states}(\omega\sin\theta_{\rm kB}, B)} at the point of pair
creation in the attenuation length calculation described in
Section~\ref{sec:attenl}.  Photons were injected at the stellar surface
with momenta along the local field lines and followed along their
curved trajectories through the magnetosphere.  For very low photon
energies \teq{\omega\sin\theta_{\rm kB} < 1+(1 + 2j_{\rm max}B/B_{\rm
cr})^{1/2}}, \teq{N_{\rm states} (\omega\sin\theta_{\rm kB}, B)} is
sampled from a table constructed by summing the exact number of pair
states (i.e. threshold energy values \teq{\omega_{j,k}} exceeded) as a
function of \teq{\omega\sin\theta_{\rm kB}}.  At higher photon
energies, \teq{N_{\rm states}(\omega\sin\theta_{\rm kB}, B)} is
computed directly from equation (\ref{eq:Nstates}) for the particular
values of \teq{\omega\sin\theta_{\rm kB}} that are sampled at the point
of pair production.

\vspace{-0.35in}
\figureout{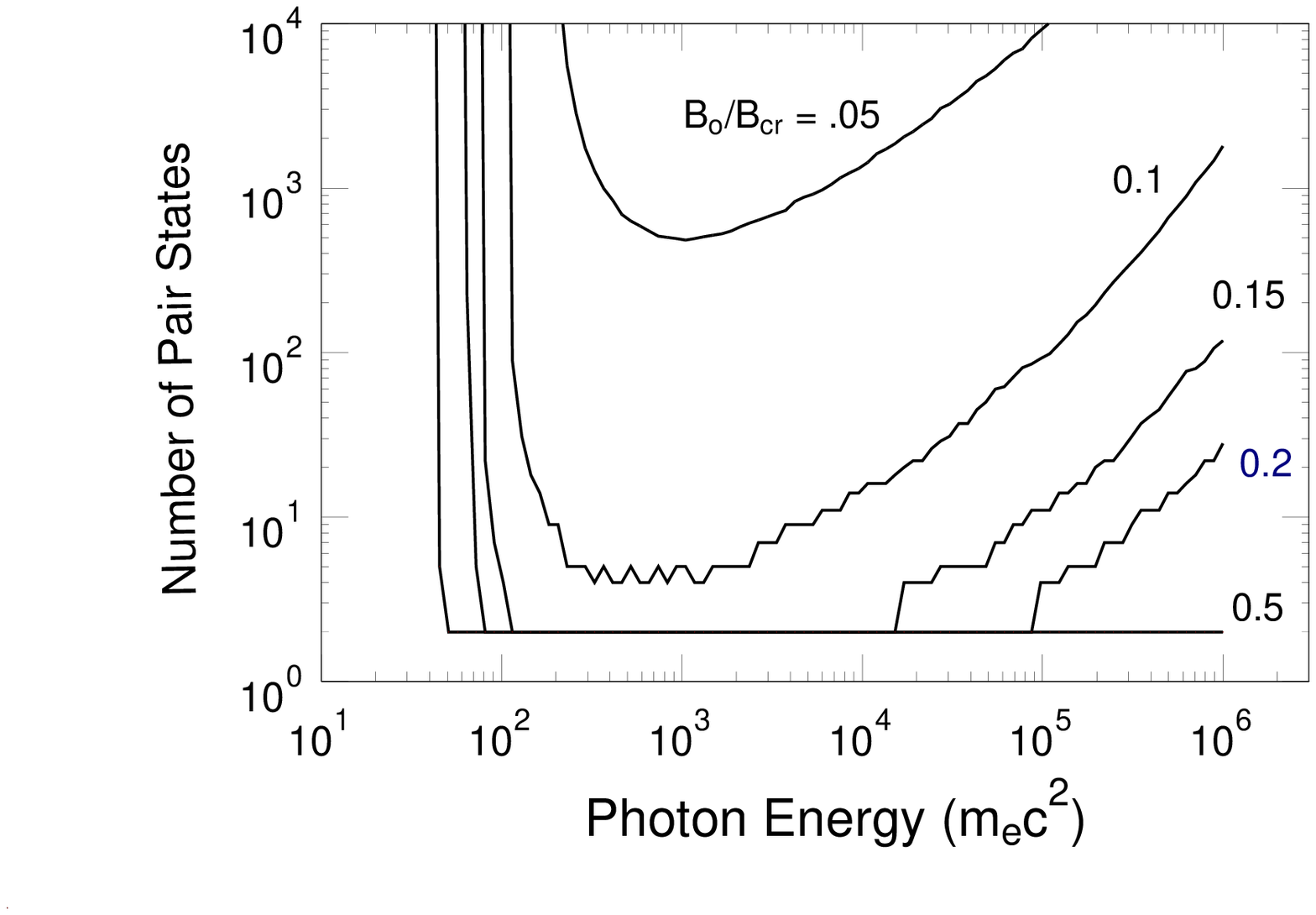}{4.6}{0.0}{0.0}{
The number of kinematically-accessible electron-positron pair states
 \teq{N_{\rm states}}, at the point of pair creation, of a photon having
 polarization state \teq{\perp}, emitted from the neutron star surface,
 parallel to the magnetic field at colatitude \teq{\Theta = 5^\circ}.
 \teq{N_{\rm states}} is displayed as a function of the photon energy in
 the observer's frame at infinity.  The curves are labeled with values
 of the dipole spin-down magnetic field strength \teq{B_0}.
 Single-state pair creation becomes rife when \teq{B_0\gtrsim 0.15B_{\rm cr}}.
 The vertical asymptotes to the left mark the respective escape energies.
 \label{fig:pairstates} }      

Figure~\ref{fig:pairstates} shows the number of available pair states
at the point of pair production as a function of the lab-frame energy
of a photon emitted parallel to the magnetic field at colatitude
\teq{\theta = 5^\circ}.  Essentially, this Figure is a modified version
of Figure~2 of Harding \& Daugherty (1983), which treated only flat
spacetime and did not examine behaviour near the escape energies (i.e.
regimes where field decline with radius plays a role).  The curves in
Figure~\ref{fig:pairstates} all increase sharply at low photon
energies, near the escape energy, where the attenuation length goes to
infinity.  At the escape energy, \teq{N_{\rm states}(\omega_{_{\rm
CM}}, B)} formally goes to infinity because photons at this energy have
large attenuation lengths and thus reach the outer magnetosphere where
the field strength has decreased.  At high energies, pair creation
accesses many states even though it occurs in the inner magnetosphere,
and the limit \teq{\omega\sin\theta_{\rm kB} \gg 1} is realized.  In
such cases, the attenuation length is \teq{L\propto \rho_c \theta_{\rm
kB}^2}, since the path length \teq{s} scales as \teq{s\approx \rho_c
\theta_{\rm kB}} for radius of curvature \teq{\rho_c}
(\teq{\propto\Theta^{-1}}), and the function \teq{T^{\rm
pp}_{\perp}(\omega\,\sin\theta_{\rm kB} )} acts like a step function in
the spatial integration determining \teq{L}.  Hence \teq{\theta_{\rm
kB}\propto \rho_c^{-1/2}} and the number of accessible pair states
asymptotically approaches \teq{N_{\rm states} \propto \erg^4
B_0^{-2}\rho_c^{-2} \propto \erg^4 \Theta^2/B_0^2}.  Therefore, an
increase in the field \teq{B_0} or a decrease in the polar cap
colatitude \teq{\Theta} reduces the number of accessible pair states.
Because the pair attenuation coefficient increases exponentially as
\teq{\exp(-4B_{\rm cr}/[3\omega B\sin\theta_{\rm kB}])}, pair
production occurs well above threshold in relatively low surface fields
(\teq{B_0 \ll 0.1B_{\rm cr}}), and photons of all energies create pairs
in highly excited states.  These pairs then produce copious numbers of
synchrotron photons, which initiate pair cascades with multiplicities
of \teq{\sim 10^3 - 10^4} (e.g. Daugherty \& Harding 1982).  In higher
fields (\teq{B_0 > 0.1B_{\rm cr}}), some photons in the spectrum create
pairs in the ground state (0,1) or (1,0) for $\perp$ polarization.  By
\teq{B_0 = 0.5B_{\rm cr}}, virtually the entire spectrum of photons
will create pairs in state (0,1) or (1,0).  The ``noisiness'' at the
minimum of the $B_0 = 0.1$ curve in Figure~\ref{fig:pairstates} is a
real discreteness effect, due to the fact that the field at the pair
creation point is increasing with photon energy and the ordering of the
(j,k) states depends on the local field strength. From these results,
we can conclude that there will be a decrease in pair yield in pulsar
polar cap cascades at magnetic fields \teq{B_0 > 0.1B_{\rm cr}}.  For
\teq{B_0 > 0.2B_{\rm cr}}, there will be strong suppression of cascade
pair yields.  Note that this condition relates to surface polar
fields.  The site for pair creation is generally proximate to the pole
only for photon energies exceeding around 1 GeV in
Figure~\ref{fig:pairstates}; otherwise the local field drops below
\teq{B_0} as the escape energy is approached.

As mentioned above, inverse-Compton scattering of pairs in the ground
state could excite them to higher Landau states, thereby increasing the
number of synchrotron photons radiated and subsequently the pair
yields.  In the case of highly relativistic particles moving along the
magnetic field, the incident photons have angles to the field near zero
in the particle rest frame.  For such photon incidence angles,
the scattering cross section in a magnetic field is
strongly suppressed below the cyclotron resonance, \teq{\ecyc =
B/B_{\rm cr}}.  Above the resonance, the particle may be excited to a
higher Landau state.  Although the scattering cross section does not
have resonances at photon energies above the cyclotron fundamental for
incidence along the field, excitation through continuum scattering at
these energies is still possible, and becomes more probable with
increasing field strength (Daugherty \& Harding 1986; Gonthier et al.
2000).  If the incident soft photons have a quasi-isotropic blackbody
spectrum at temperature \teq{T} and their average energy is
\teq{\omega_s = 3kT}, then scattering above the resonance in the
particle rest frame will occur when \teq{\gamma \gtrsim (B/B_{\rm
cr})\, m_ec^2/3kT = 2 \times 10^3\,(B/B_{\rm cr})\,T_6^{-1}},
a condition applicable for all fields provided that the particles are
fairly relativistic.  The mean-free path of these particles to
scattering will then be
\begin{equation}
   \lambda_s \simeq {1\over \eta\sigma_T n_{\gamma}}\; =\;
   7.5 \times 10^4\,\eta^{-1}\,T_6^{-3}\, \rm cm,
 \label{eq:lams}
\end{equation} 
where \teq{T_6 \equiv T/10^6} K and \teq{\eta} is a suppression factor
for the scattering cross section \teq{\sigma = \eta\sigma_T} in the
Klein-Nishina regime.  Such mean free paths are short enough to suggest
that excitation through scattering is potentially important.  However,
these photons scattering above the resonance in the electron rest frame 
{\it will not form the bulk of the produced pairs}, because the resonant
cross section (and thus the scattering rate) is several orders of 
magnitude larger than for non-resonant cross section.

\subsection{Positronium Formation}
 \label{sec:suppposit}

Another mode of pair creation exists, namely the formation of pairs in
a bound state, i.e. positronium.  This has been proposed as an
effective competitor to the production of free pairs (Shabad \& Usov
1985, 1986; Herold, Ruder \& Wunner 1985; Usov \& Shabad 1985; Usov \&
Melrose 1995) because the binding energy lowers the threshold slightly
(\teq{\ll 1}\%) below the value for production of free pairs.  Positronium
formation fundamentally alters the magnetosphere:  it amounts to the 
suppression of electron or positron
currents that can screen the induced electric fields in the rotator.
Furthermore, depending on how long the pairs remain bound in their passage 
along open field lines to the light cylinder, such a presence of neutral
positronium will help to suppress any collective plasma modes that
might spawn radio emission.  Hence positronium formation is a viable
alternative means of inhibiting the radio signal from pulsars.

The relevance of positronium formation to pulsars is potentially great
if the bound state is stable for considerable times.  Positronium is
subject to destruction by three main mechanisms: free decay, electric
field ionization,  and photo-ionization.  Using the spontaneous
two-photon decay rate (in the positronium rest frame) of \teq{3.5\times
10^{14}\, (B/B_{\rm cr}) \hbox{sec}^{-1}} computed by Wunner \& Herold
(1979), Bhatia, Chopra \& Panchapakesan (1987) obtained positronium
decay lengths considerably shorter (by two orders of magnitude for bulk
Lorentz factor \teq{\gamma_{\rm p} =10^3}) than the stellar radius for
\teq{B\sim 0.2B_{\rm cr}}; the length is a strongly declining function
of \teq{B}.  Hence it appears that positronium is unlikely to be
long-lived in the magnetosphere except for a narrow range of fields, or
for sufficiently high positronium Lorentz factors, i.e. from the
attenuation of photons above 10 GeV.  Approximate estimates of when
field ionization becomes important were obtained by Herold, Ruder \&
Wunner (1985) and Usov \& Melrose (1996): for pulsar periods
\teq{P\lesssim 0.5 (10B_0/B_{\rm cr})^{2/3}}sec, ionization due to
\teq{E_\parallel} should convert positronium into free pairs.  This
corresponds to a sizeable portion of the phase space of interest for
the considerations of this paper.  In addition, photo-ionization can
destroy positronium, forming free pairs.  Herold, Ruder \& Wunner
(1985) claimed that this was highly likely due to collisions with
thermal radiation emanating from the stellar surface for temperatures
greater than around \teq{10^5}K.  Usov \& Melrose (1995) used the
calculations of Bhatia, Chopra \& Panchapakesan (1992) to argue that
photo-ionization timescales exceed dynamical ones for surface
temperatures \teq{\sim 10^5}K.  This discrepancy is yet to be
resolved.

\subsection{Pair Cascades in High Magnetic Fields}
 \label{sec:cascades} 

We have discussed several of the physical mechanisms capable of
suppressing pair production when the local magnetic fields approach and
exceed the critical field.  To more realistically measure pair
suppression in high magnetic fields, one must examine not only the
propagation of single high-energy photons through the neutron star
magnetosphere (i.e., the escape energies) but also the transport of the
higher generations of photons and particles, which are produced by the
primary photons that do not escape.  Even though the first generation
of photons split instead of producing pairs, the second generation of
photons may create pairs.  It is therefore necessary to investigate
pair yields of the pair/photon cascades which are initiated by the
primary photons.  In Section~\ref{sec:simcas}, we will present results
of numerical simulations of such pair cascades.  In this section, we
will qualitatively discuss how we expect the nature of these cascades
to change as the magnetic field increases.

\vspace{0.25in}
\figureout{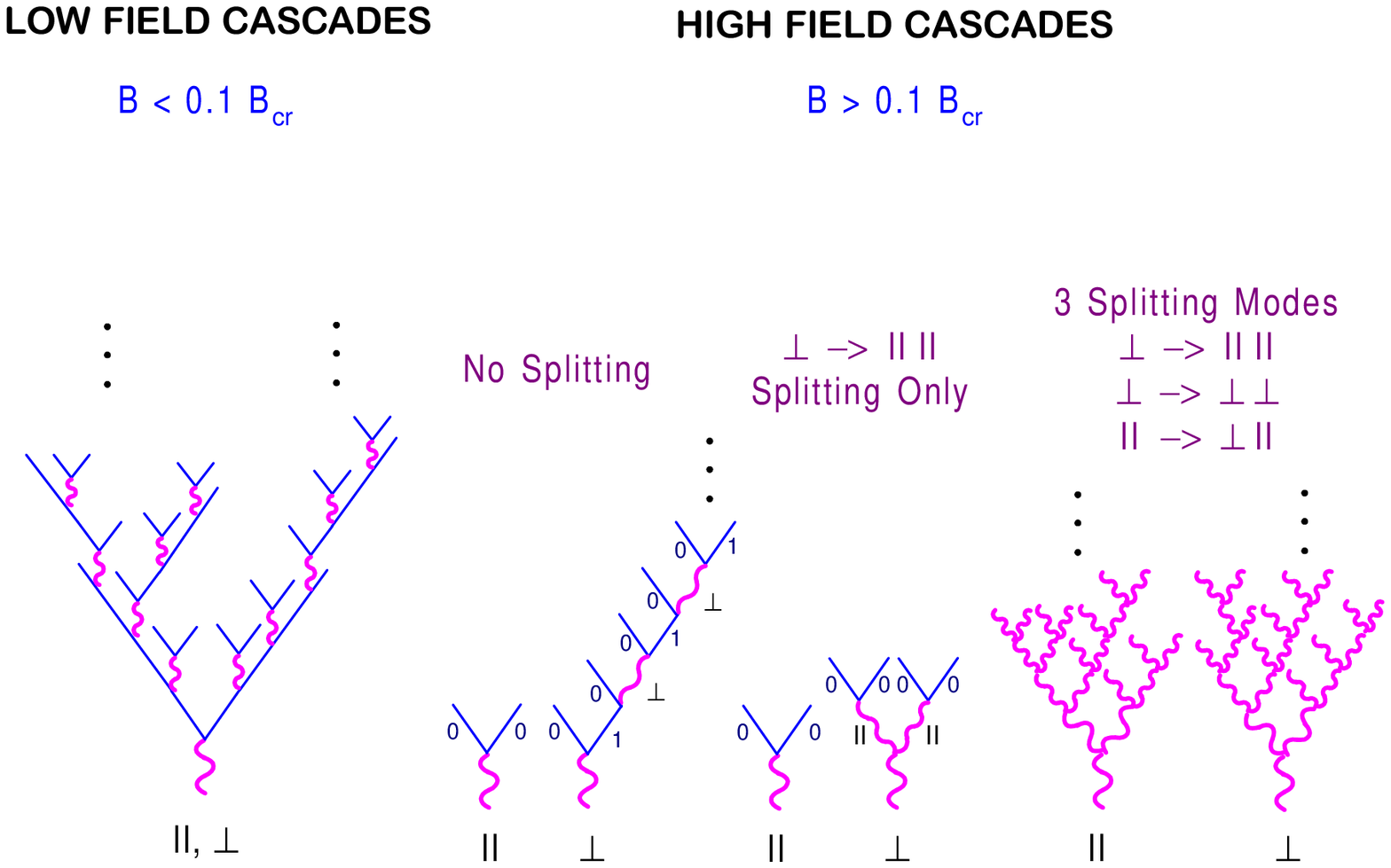}{3.4}{0.0}{0.0}{
The nature of pair cascades in increasing magnetic fields.  Wavey lines
represent photons and straight lines represent electrons and
positrons.  Polarization modes of photons (\teq{\parallel} - parallel or
\teq{\perp} - perpendicular) are noted where relevant.
 \label{fig:paircas}}           

Figure~\ref{fig:paircas} illustrates the behavior of low and high field
pair cascades.  Pair cascades in magnetic fields well below \teq{B_{\rm
cr}} have been extensively studied in connection with models of pulsar
high energy emission (e.g. Daugherty \& Harding 1982, 1996).  In fields
\teq{B \lesssim 0.1 B_{\rm cr}}, both \teq{\perp} and \teq{\parallel}
mode photons above escape energy will produce pairs well above
threshold, so that the resulting electrons and positrons will initially
occupy high Landau states.  Each pair member will make multiple Landau
state transitions before reaching the ground state, radiating many
synchrotron photons.  Many of the synchrotron photons will pair
produce, initiating more synchrotron emission sequences.  Such a
process can go on for a number of generations, provided that the energy
of the primary photon is high enough, resulting in the production of
many pairs for each primary photon.

When \teq{B \gtrsim 0.1 B_{\rm cr}}, the primary photons create pairs at
pair threshold, as discussed in Section~\ref{sec:suppground}.  This has
a profound effect on the pair cascades, since the \teq{\parallel} mode
photons produce pairs with both members in the ground state, cutting off
all further generations.  Photons of \teq{\perp} mode, which have a higher
threshold, produce pairs with one member in the first excited state.
The single Landau state transition made by this particle will result in
primarily \teq{\perp} mode photons which may create a second generation
pair with one member in the first excited state.  Higher generations of
pairs are therefore possible, depending on the energy of the primary
photon, but the total pair yield is severely diminished because there
is no multiplication of pairs with each generation, as in the low-field
cascades.

At fields \teq{B \gtrsim 0.5 B_{\rm cr}}, the photon splitting attenuation
lengths become shorter than pair attenuation lengths.  At this point,
the nature of the pair cascades will depend on which modes of splitting
are operating at what field strengths.  In fields \teq{B \lesssim B_{\rm
cr}}, it is probable that only the \teq{\perp \rightarrow
\parallel\parallel} mode operates (see Section~\ref{sec:psplit}).
Thus, primary \teq{\parallel} mode photons can produce pairs, but with both
members in the ground state.  The primary \teq{\perp} mode photons above
photon splitting escape energy will split into two \teq{\parallel} mode
photons, each of which can produce a pair in the ground state.  Thus,
each polarization branch of the cascade ends after, at most, one pair
generation.  The action of photon splitting in one mode therefore will
lower the cascade pair yield above \teq{0.5 B_{\rm cr}}, but only
moderately.

If, as was discussed in Section~\ref{sec:suppsplit}, other modes of
photon splitting open up in magnetic fields above \teq{B_{\rm cr}} due
to increased vacuum dispersion, then the character of pair cascades
undergo a further transition.  If both photon polarization modes can
split (as is the case when all three modes permitted by QED are
operating), then it is possible to completely prevent the production of
pairs.  A pure photon splitting cascade then ensues, in which photons
split repeatedly until they can escape the magnetosphere.  The only
question then becomes whether the splitting cascade can degrade the
photon energy fast enough so that the photons escape without pair
conversions in lower fields.  Since the photons are splitting below pair
threshold in high fields, this would seem very likely.

\section{PAIR SUPPRESSION IN HIGHLY-MAGNETIZED PULSARS}
 \label{sec:noradio}

Pair creation is obviously pertinent to the hard X-ray and gamma-ray
emission of pulsars.  Yet it is also relevant to the discussion of
coherent emission in radio pulsars, since it is commonly assumed that a
plentiful supply of pairs is a prerequisite for, and maybe also a
guarantee of, coherent radio emission at observable flux levels.  Such
a connection is the premise of standard models for radio pulsars (e.g.
Sturrock 1971; Ruderman \& Sutherland 1975; Arons \& Scharlemann 1979),
a relationship founded in the simplicity of its explanation (Sturrock,
Baker \& Turk 1976) for the extinction of radio pulsars beyond the
conventional death line at long periods.  Detailed discussions of the
relationship of pairs to the production of radio emission can be found
in Michel (1991) and Melrose (1993), though we note that there are
dissenting views to this popular connection (e.g. Weatherall \& Eilek 1997).

The pre-eminent consequence of any suppression of pair creation in
pulsars is that the emission of radio waves should be strongly
inhibited.  Hence the issue of possible quenching at high \teq{B_0} by
the mechanisms discussed above forms the focus of this Section.  First
we compute the putative radio quiescence boundaries in the \teq{\dot
P}--\teq{P} diagram using the approximate criterion for the suppression
of pair creation by photon splitting defined in
Section~\ref{sec:erg_escape}, namely when the escape energy for the
splitting of \teq{\perp} photons is less than or equal to that for pair
creation \teq{\perp\to e^{\pm}}.  Essentially this provides a detailed
derivation of the boundary introduced in Baring \& Harding (1998).  We
then model the physics of pair suppression near pulsar polar caps by
means of a Monte Carlo simulation of pair cascades in high fields,
treating both photon polarization states.  These detailed calculations
identify the requirements for effective pair suppression which must be
met to justify the simplified criterion for pair suppression of Section
\ref{sec:erg_escape}.

\vspace{0.25in}

\subsection{Radio Quiescence Boundaries in the \teq{P}-\teq{\dot{P}} Diagram}
 \label{sec:quiescence}

In Section~\ref{sec:erg_escape}, the condition for the equality of
photon splitting and pair creation escape energies established a region
in \teq{(B_0,\,\Theta )} space where pair suppression would probably
occur.  This space can be easily transformed into a region in the radio
astronomer's traditional pulsar phase space, which consists of the
measurables \teq{P}, the pulsar period, and \teq{\dot{P}}, the period
derivative.  The first part of this transformation derives from the
relationship between the period and the size of the polar cap
\teq{\Theta}.  Such a relation follows from the definition of the cap
as the portion of the stellar surface that anchors open field lines,
i.e. those that are not closed within the light cylinder.  In flat
spacetime, for a rotating dipole field, the angle \teq{2\Theta}
subtended by the polar cap can be expressed via (e.g. Manchester and
Taylor 1977) \teq{\sin\Theta =[2\pi R/(P\, c)]^{1/2}}.  Due to general
relativistic distortions of the dipole in a Schwarzschild metric, this
formula is modified to the form (Muslimov and Harding 1997)
\begin{eqnarray}
   \sin\Theta &=& \Biggl\{ \dover{2\pi R}{P c}\; 
   {\cal F}\biggl(\dover{R_s}{R}\biggr)\Biggr\}^{1/2}\;\; ,
 \nonumber\\[-5.5pt]
 \label{eq:polarcap}\\[-5.5pt]
 {\cal F}(x) &=& -\dover{x^3}{3}\,\biggl[ \log_e(1-x) + 
   \dover{x}{2}\, (x+2) \biggr]^{-1} \nonumber
\end{eqnarray}
that is used in this paper.  Here \teq{R_s=2GM/c^2} is the
Schwarzschild radius, and the result for any radius \teq{r} can be
obtained by the substitution \teq{R\to r}.  Clearly, the strong
gravitational field reduces the size of the polar cap (\teq{{\cal
F}(x)\approx 1-3x/4} for \teq{x\ll 1}), arising in conjunction with its
intensification of the magnetic field.  Note that strictly speaking,
the polar cap connects to field lines that are not closed within the
Alfv\'en radius \teq{r_{\hbox{\sixrm A}}}, however, the intense pulsar
fields yield relativistic Alfv\'en speeds so that \teq{r_{\hbox{\fiverm
A}}} is generally close to the light cylinder radius \teq{P\, c/(2\pi
)}.  The loading of the field with plasma can have a significant impact
on \teq{r_{\hbox{\sixrm A}}} and \teq{\Theta} in soft gamma repeaters, 
as discussed in Section~\ref{sec:magnetars} below.

The second half of the transformation of the \teq{(B_0,\,\Theta )}
space to the (\teq{P,\, \dot{P}}) phase space arises from the commonly
assumed spin-down relationship between the dipole spin-down field
\teq{B_0} and the measurables.  For radio pulsars, the increase in
period is usually attributed to magnetic dipole radiation that taps the
rotational kinetic energy of the neutron star.  This couples \teq{B_0}
to \teq{P} and \teq{\dot{P}} through equating the rotational energy
loss to the dipole radiation energy loss which is specified in terms of
the total magnetic moment \teq{\mu_{\hbox{\fiverm B}}}.  The radio
pulsar community has commonly adopted the approximation
\teq{\mu_{\hbox{\fiverm B}} \sim B_0R^3} (e.g. see Manchester and
Taylor 1977).  The actual dipole moment of a star-centered dipole,
regardless of internal field configuration, is \teq{\mu_{\hbox{\fiverm
B}}=(B_0R^3)/2} (e.g. see Shapiro and Teukolsky 1983, Usov and Melrose
1995).  This choice, which we adopt throughout this paper, leads to a
dipole radiation loss rate of \teq{dE/dt =-B_0^2\, R^6\Omega^4/(6c^3)}
for an orthogonal rotator of angular frequency \teq{\Omega =2\pi/P},
and
\begin{equation}
   B_0\; =\; 6.4\times 10^{19}\,\sqrt{P\, \dot{P}}\;\;\hbox{Gauss}
 \label{eq:spindown}
\end{equation}
for the surface field at the magnetic pole.  This estimate is twice the
conventional choice of Manchester and Taylor (1977).  We note that
multipole contributions near the stellar surface and the field geometry
in the outer magnetosphere can also influence the pulsar surface fields 
derived from \teq{P} and \teq{\dot{P}}.  The uncertainties in
the understanding of the global field structure are sufficient that the
use of Eq.~(\ref{eq:spindown}) is justified for the purposes of this
paper.  Note that curved spacetime does not impact the determination of
\teq{B_0}, a flat spacetime quantity, but just increases the value of
the field in the local frame at the pole over \teq{B_0}; this increase
in magnetic field energy density can be viewed as a general
relativistic ``redshifting'' of the Poynting flux of the rotator.

\vspace{0.1in}
\figureout{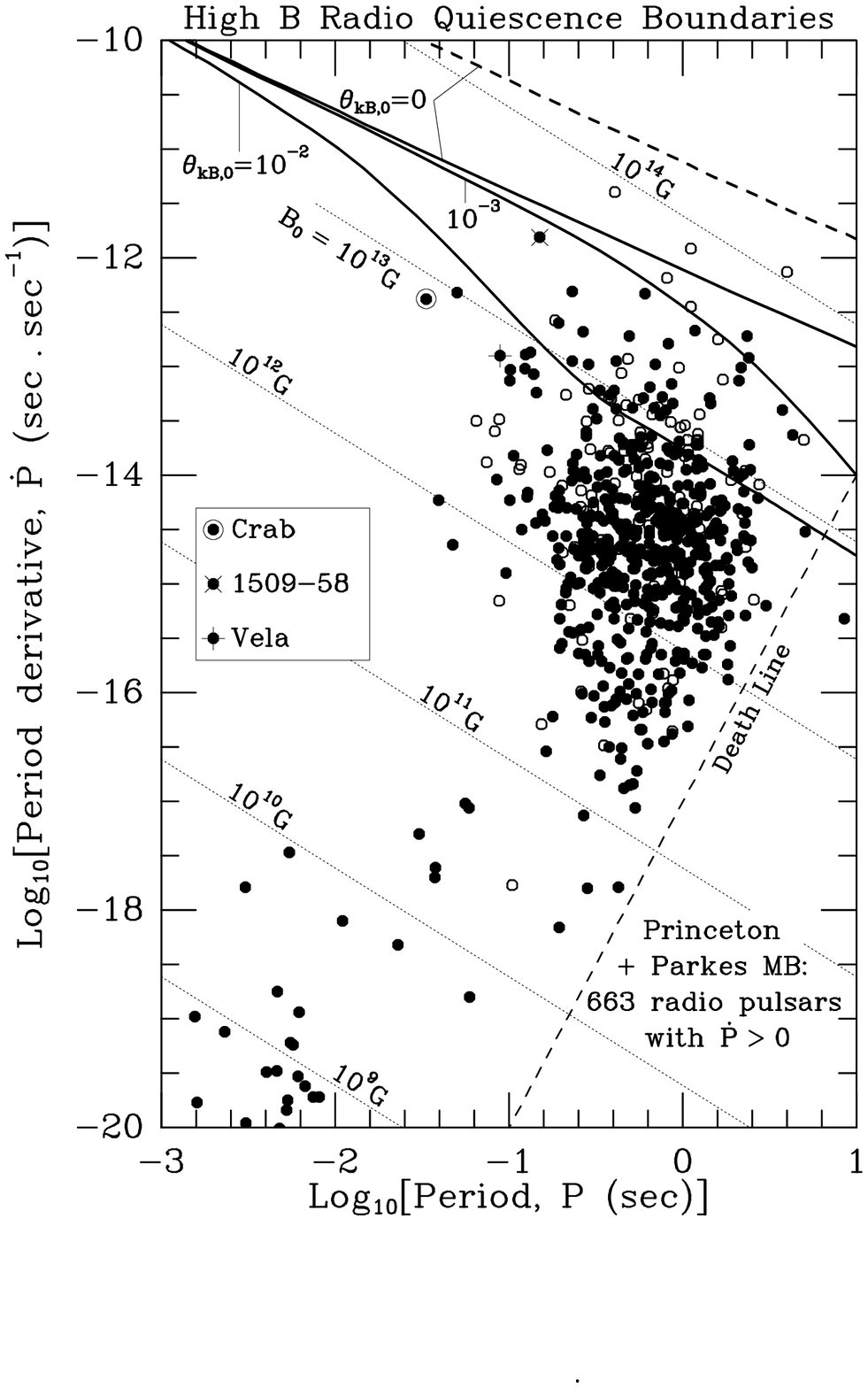}{3.75}{-0.15}{-0.82}{
The conventional depiction of pulsar phase space, the
 \teq{P}-\teq{\dot{P}} diagram, with filled circles denoting the
 locations of 541 members (those with \teq{{\dot P}>0}) of the latest
 edition (Taylor, Manchester \& Lyne 1995; see also {\tt
 http://pulsar.princeton.edu/}) of the Princeton Pulsar Catalogue, and
 open circles marking 122 pulsars in the recent Parkes Multi-Beam survey
 [{\tt http://www.atnf.csiro.au/\~{}pulsar/psr/pmsurv/pmwww/}].  Three
 of the handful of gamma-ray pulsars, namely the Crab, Vela and PSR
 1509-58 are highlighted, with point styles as indicated in the inset.
 The dotted diagonal lines denote constant field strength, as inferred
 from Eq.~(\ref{eq:spindown}).  A fiducial positioning of the
 conventional death line, from \teq{P}-\teq{\dot{P}} diagrams at {\tt
 http://pulsar.princeton.edu/}, is depicted as the dashed line on the
 right.  The heavy solid curves give a variety of choices for the
 boundary of radio quiescence for surface emission, depending on the
 value of \teq{\theta_{\rm kB,0}}, the initial angle of photons relative
 to the field.  In addition, the heavy dashed curve is a similar
 boundary for \teq{\theta_{\rm kB,0}=0} and emission point half a
 stellar radius above the surface (i.e. \teq{R_0=1.5R}).  Each of these
 curves, defined by solutions to
 Eqs.~(\ref{eq:ergescsupp})--(\ref{eq:spindown}), purportedly partitions
 the \teq{P}-\teq{\dot{P}} diagram into regions of radio loud (below the
 curve) and radio quiet (above) pulsars, the latter being where photon
 splitting suppresses pair creation for the \teq{\perp} polarization
 state.  Values \teq{\theta_{\rm kB,0}\lesssim 10^{-4}} and
 \teq{r\gtrsim 1.2R} can comfortably accommodate the current radio
 pulsar population.
\label{fig:quiescence} }      

By inverting Eq.~(\ref{eq:spindown}) to solve for \teq{\dot{P}} in
terms of \teq{B_0} and \teq{P}, the solutions \teq{B_0(\Theta )} to
Eq.~(\ref{eq:ergescsupp}) that demarcate the domain of possibly strong
suppression of pair creation by photon splitting are mapped onto
the \teq{P}-\teq{\dot{P}} diagram in Figure~\ref{fig:quiescence}.
Curves are depicted for three of the four values of \teq{\theta_{\rm
kB,0}} that are addressed in Fig.~\ref{fig:ergescsol}, specifically for
the case of surface photon emission, and also for a \teq{\theta_{\rm
kB,0}=0} case of emission above the surface.  Much of the extant isolated
radio pulsar population is also exhibited (only those with positive
\teq{\dot P}), namely those sources listed in the Princeton Pulsar
Catalogue (Taylor, Manchester \& Lyne 1993, with 541 pulsars: see {\tt
http://pulsar.princeton.edu/}), with a recent update from the Parkes
Multi-Beam Survey (adding 122 new pulsars with presently archived
\teq{P} and \teq{\dot P}: see {\tt
http://www.atnf.csiro.au/\~{}pulsar/psr/pmsurv/pmwww/}).  Note that,
from the discussion in Section~\ref{sec:erg_escape}, the
\teq{\theta_{\rm kB,0}=0} boundary for surface emission (\teq{R_0=R})
corresponds roughly to \teq{B_0\propto \Theta^{-4/15}} and hence to
\teq{\dot{P}\propto P^{-11/15}}; specifically, we find the
approximation
\begin{equation}
   \dot{P}\;\approx\; 7.9\times 10^{-13}\; 
   \biggl(\dover{P}{1\,\hbox{sec}}\biggr)^{-11/15}\quad ,
 \label{eq:deathline}
\end{equation}
the slope of which depends on the how strongly the rate of photon
splitting depends on \teq{B} in this regime of fields. It should also
be noted that while such putative {\it boundaries of radio quiescence}
have been labelled by specific values of \teq{\theta_{\rm kB,0}} for
the purpose of illustration, in fact the value of this initial angle of
photon propagation with respect to the field is a weak function of the
colatitude.  This is due to the coupling between the Lorentz factor of
accelerated electrons (and therefore \teq{\theta_{\rm kB,0}}) and the
polar cap size \teq{\Theta}, a connection discussed below in
Section~\ref{sec:accel}.  Also depicted in Figure~\ref{fig:quiescence}
is a fiducial positioning (adopted by Taylor, Manchester \& Lyne 1993)
of the conventional death line, with \teq{\dot{P} \propto P^{3}} (i.e.
corresponding fixed open field line voltage, adopt), to the right of
which there is just one detected radio pulsar (PSR J2144-3933).  It is
a well-known problem that the computed position of the pulsar ``death
line", assuming pairs are produced only by curvature radiation photons,
lies at smaller periods than the observed ``death line", for the
assumption of the standard polar cap size (e.g.  Arons \& Scharlemann
1979).  However, increasing the polar cap size to less than twice that
of the standard brings the computed ``death line" into agreement with
virtually the entire radio pulsar population, the notable exception
being the recently ``re-discovered'' 8.5 second Parkes pulsar (Young,
Manchester \& Johnston 1999) that is conspicuous on the right hand side
of Fig.~\ref{fig:quiescence}.  However, pair production by inverse
Compton-scattered photons can explain radio emission from this pulsar
(Zhang et al. 2000).

The Princeton catalog by itself clearly rules out boundaries of radio
quiescence due to photon splitting that correspond to \teq{\theta_{\rm
kB,0}\gtrsim 3\times 10^{-4}} for surface emission.   The fact that the
\teq{\theta_{\rm kB,0}=0}, \teq{R_0=R} boundary was comfortably located
above the entire Princeton collection of radio pulsars was an
attractive feature that underpinned the suggestion of Baring \& Harding
(1998) that this particular case marked the approximate location of the
boundary of radio quiescence, where photon splitting could strongly
inhibit pair creation.  However, this conclusion has been challenged by
the dramatic increase in the observed population due to the exciting
new Parkes Multi-Beam survey (Camilo et al. 2000; D'Amico et al. 2000;
Kaspi et al. 2000), with three new pulsars (PSRs J1119-6127, 1726-3530
and 1814-1744, all with high dispersion measures in the 700--850 range)
having been discovered that are obviously at \teq{\dot P} above the
\teq{\theta_{\rm kB,0}=0}, \teq{R_0=R} boundary. 

\clearpage

Although the boundary of radio quiescence for surface emission (chosen
for illustrative purposes by Baring \& Harding 1998) neatly falls
between the population in the Princeton catalog and the handful of
known anomalous X-ray pulsars and soft gamma repeaters (see the blow-up
of the \teq{P}--\teq{\dot P} diagram in Figure~\ref{fig:escape_cont}),
most (or perhaps all) of which are radio quiet, clearly the new Parkes
survey data questions this assumption of surface emission.  The
quiescence boundary rapidly moves up on the \teq{P}-\teq{\dot{P}}
diagram for emission at higher altitudes, which might naturally be
expected for traditional curvature radiation-initiated cascades in
pulsars with Crab-like surface fields (e.g. around \teq{R} above the
surface for Daugherty \& Harding's 1996 modeling of the Vela light
curve).  However, for the high \teq{B} fields sampled by the
\teq{\theta_{\rm kB,0}\lesssim 10^{-4}} curves, inverse Compton is
probably more relevant to the acceleration region (see Sturner 1995;
Harding \& Muslimov 1998) and low altitude emission is much more
probable, as discussed in Section~\ref{sec:accel} below.  One might
then expect that the mean radius of emission satisfies \teq{r\lesssim
1.5R}.  Accordingly, we have computed a \teq{\theta_{\rm kB,0}=0},
\teq{R_0=1.5R} boundary and depicted it in Figure~\ref{fig:quiescence};
it clearly lies above even the highly-magnetized Parkes Multi-Beam
survey pulsars.  Intuitively, it might be anticipated that the location
of the boundary as a function of emission radius $R_0$ should scale
simply by the relationship between field strength and radius for a
dipolar structure, so that \teq{B_0\propto r^3}.  This appears to be
borne out near the stellar surface, however the radial dependence of
the location of the quiescence boundary weakens at altitudes \teq{r\sim
R} (in spite of the reduced effects of general relativity at
\teq{r>R}), principally because of changes in the flaring of field
lines sampled by photons in their flight away from the neutron star.
The present Parkes survey data constrain computed quiescence boundaries
to the \teq{r\gtrsim 1.2R} range; in the light of the present
indeterminacy in the actual locale of the acceleration region
(addressed in Section~\ref{sec:accel}), this is not a serious problem.

\subsection{Simulations of Photon Splitting/Pair Cascades}
\label{sec:simcas}

Using the equality of photon splitting and pair creation escape
energies for \teq{\perp} polarization as the criterion for when
splitting starts to strongly suppress the production of pairs is a
simple choice, motivated largely by the predominance of production of
\teq{\perp} photons in the relevant emission processes.  It is
necessary to determine whether and under what conditions this choice is
an appropriate one. 

To assess the applicability of the radio quiescence boundaries, we have
studied pair suppression in high magnetic fields by means of Monte
Carlo simulations of photon splitting/pair cascades in a neutron star
magnetosphere.  This calculation generalizes that of HBG97 by including
the cyclotron and synchrotron radiation from the electron-positron
pairs.  This improvement is necessary in order to compute the pair
yields from all generations of the cascade.  Note that we omit resonant
Compton scattering from our simulations since, while it is important as
a radiation process for primary and secondary particles, it will not
significantly impact the spatial transfer of photons in normal radio
pulsars due to the low densities and small angles; it may, however, be
significant in soft gamma repeaters.  We inject photons parallel to the
local magnetic fields at the neutron star surface with specified
magnetic colatitude \teq{\Theta} and four-momentum \teq{k}.  The photon
energies are sampled from a power-law distribution,
\begin{equation} 
   N(\erg) = N_0\erg^{-\alpha},  \quad
     \erg_{\rm min} < \erg < \erg_{\rm max}\; ,
  \label{eq:N}
\end{equation}
where we generally set \teq{\erg_{\rm min}=1}, and let \teq{\erg_{\rm
max}} be a free parameter.  In the results presented in this paper, we
generally take $\alpha = 1.6$ as a spectral index representative of
either primary curvature radiation with energy losses (giving $\alpha =
5/3$) or resonant inverse Compton emission which is relatively hard.
Polarization is chosen randomly such that the average distribution has
\teq{75\%} in the \teq{\perp} mode and \teq{25\%} in the
\teq{\parallel} mode, reflecting the polarization produced by the
radiation mechanisms discussed at the beginning of
Section~\ref{sec:suppression}.  The path of each input photon is traced
through the magnetic field, in curved spacetime, accumulating the
survival probabilities for splitting, \teq{P_{\rm surv}^s}, and for
pair production, \teq{P_{\rm surv}^p}, independently:
\begin{equation}
   P_{\rm surv}(s) = \exp\Bigl\{-\tau(\Theta, \erg,\, l)\Bigr\} 
\end{equation}
where \teq{\tau(\Theta, \erg,\, l)} is the optical depth as defined in
Eqn.~(\ref{eq:tau}).  Each photon may split,
produce pairs or escape, based on a combination of the running survival
probabilities for splitting and pair production (see HBG97).  If a
photon splits, the energies and polarizations of the final photons are
sampled from the distribution from the branching ratios given in
HBG97.  Each final photon is then followed in the same way as the
parent photon.  If a photon produces pairs, the total energy, Landau
state and parallel momentum of the electron and positron are
determined.  Each member of the pair is assumed to have half the energy
and the same direction of the parent photon (technically this is only
appropriate for high Landau states), except when the pair is produced
at threshold (i.e. in the [0,0] state for \teq{\parallel} polarization
and [0,1] or [1,0] for \teq{\perp} polarization), in which case we use
the full kinematic equations to determine the energy and momentum of
each pair member.  Each member of the pair occupying an excited state
emits a sequence of cyclotron or synchrotron photons.  The method used
to simulate the cyclotron/synchrotron emission is similar to that of
Daugherty \& Harding (1996).  If the particle Landau level is larger
than 20, the high-energy limit of the quantum synchrotron transition
rate (Sokolov \& Ternov 1968) is used, in which case we assume that the
photons are emitted perpendicular to the magnetic field in the particle
rest frame (high-energy limit).  When the particle Landau level is
smaller than 20, the exact QED cyclotron transition rate (Harding \&
Preece 1987) is used, in which case the angles of the emitted photons
are sampled from a distribution.  In both cases, the emitted photon
polarizations are sampled from the corresponding polarization
distributions.  Each emitted photon is propagated through the magnetic
field from its emission point until it splits, produces pairs or
escapes.  The cyclotron/synchrotron emission sequence continues until
each particle reaches the ground state.  The cascade continues until
all photons from each branch have escaped.  Throughout the cascade, a
running tally is kept of the number of pairs produced and of the number
of photon splittings.  General relativistic effects of a Schwarzschild
metric on the photon momentum and dipole magnetic field, are included
in the same way as described in HBG97.

We have chosen the total number of pairs produced per injected photon
as a quantitative measure of the {\it pair yield} of the cascade
initiated by a particular parent photon spectrum.  The free parameters
are then surface magnetic field strength, emission colatitude
\teq{\Theta}, spectral index \teq{\alpha}, and minimum, \teq{\erg_{\rm
min}} and maximum, \teq{\erg_{\rm max}}, energy of the primary photon
spectrum.  Cascade simulations have been run for the cases where one
mode of splitting operates (the \teq{\perp \rightarrow
\parallel\parallel} mode), three modes of splitting operate (the
\teq{\perp \rightarrow \parallel\parallel}, \teq{\perp \rightarrow
\perp\perp} and \teq{\parallel \rightarrow \perp\parallel} modes
permitted by QED), and where splitting is turned off completely 
(``no splitting").
Figure~\ref{fig:py1mode} shows the cascade pair yield as a function of
magnetic field strength and maximum primary photon energy; since the
total number of photons and the pair yield depend on the value of
\teq{\erg_{\rm min}}, we hold \teq{\erg_{\rm min} = 1} constant for all
calculations.

\vspace{-0.1in}
\figureout{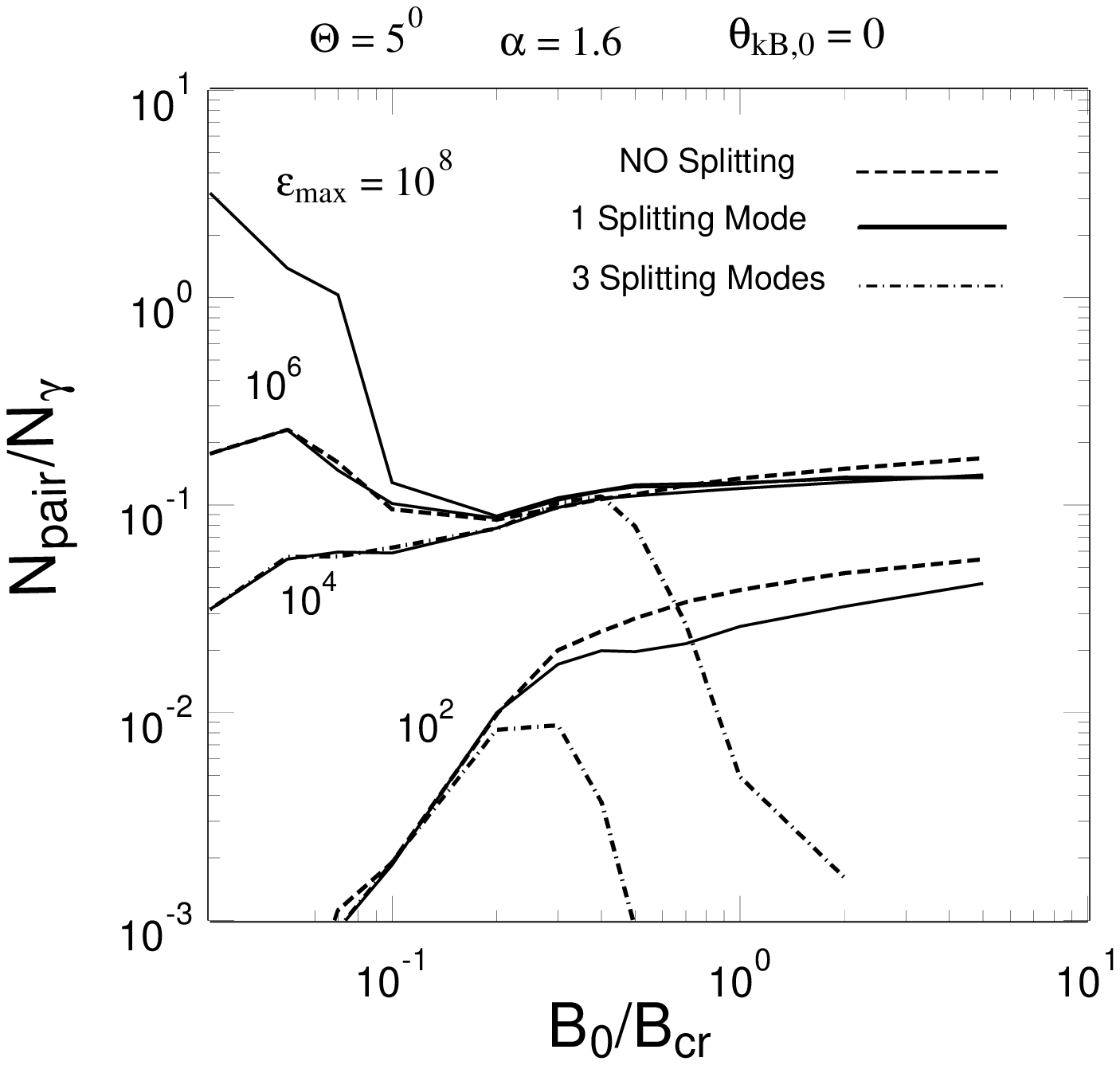}{4.9}{0.4}{-0.3}{
The cascade pair yield (number of pairs per injected photon) as a
function of surface magnetic field strength, \teq{B_0}, in units of the
critical field, \teq{B_{\rm cr}}, and maximum primary photon energy,
\teq{\erg_{\rm max}}, in units of \teq{m_ec^2}.  Different line types
refer to the cases where no splitting, only one mode
(\teq{\perp\to\parallel\parallel}) or three modes of splitting are
allowed.  The curves for the ``no splitting" case behave similarly for
\teq{\erg_{\rm max}=10^4}, \teq{\erg_{\rm max}=10^6} and \teq{\erg_{\rm max}=10^8} cases, so only the curves for the \teq{\erg_{\rm max}=10^2} and \teq{\erg_{\rm max}=10^6} cases are shown for clarity. 
Switching on only one mode of splitting leads to only minimal
reductions in the pair yield for \teq{\erg_{\max}\gtrsim 10^4}, as
explained in the text. Photons are assumed 
to be emitted from the stellar surface.
 \label{fig:py1mode}}           

The curves for the ``no splitting" case, shown only for \teq{\erg_{\rm
max} =10^2} and \teq{\erg_{\rm max}=10^6} to avoid confusion, are of
course not physical but are included to illustrate the effect of one
splitting mode on the pair yield.  When \teq{\erg_{\rm max}} is below
the pair escape energy (cf.  Section~\ref{sec:erg_escape}), as it is
for the \teq{\erg_{\rm max} = 10^2} and \teq{\erg_{\rm max} = 10^4}
curves at lower field strengths, the pair yield is considerably lower
than when \teq{\erg_{\rm max}} is well above pair escape energy (as is
the case for the \teq{\erg_{\rm max} = 10^6} and \teq{\erg_{\rm max} =
10^8} curves), for which pair yields at the lower field strengths can
be quite large due to the pair multiplication effect through successive
generations, as described in Section~\ref{sec:cascades}.  Above a field
strength of \teq{0.1 B_{\rm cr}}, in \teq{\erg_{\rm max}\gtrsim 10^4}
cases the pair yields drop because the pairs are produced in low Landau
states at or near threshold, suppressing synchrotron emission and thus
the pair multiplication effect.  In the absence of photon splitting,
the pair yield increases slowly with increasing \teq{B_0 > 0.1 B_{\rm
cr}}, because the number of generations of the cascade can increase.
The rapid rise of the \teq{\erg_{\rm max}=10^2} case when
\teq{B_0\lesssim 0.2 B_{\rm cr}} is due primarily to the escape energy
for pair production \teq{\erg_{\rm esc}^{\perp\to e^+e^-}} being close
to \teq{\erg_{\rm max}} so that the optical depth is of the order of
unity or less; the pair yield must then necessarily be a strong
function of \teq{B_0} by virtue of the exponential asymptotic form (see
Daugherty \& Harding 1983) for the pair creation rate.  Since the pair
escape energy saturates when \teq{B_0 \gtrsim 0.5 B_{\rm cr}} at
approximately the threshold value of \teq{2/\sin\Theta} (see
Figure~\ref{fig:escape}), the pair yield should asymptote to a constant
value at high field strengths.  This constant is just the fraction of
primary photons above pair threshold:  when \teq{\erg_{\max}\gg 1},
this fraction is approximately \teq{(\sin\Theta /2)^{\alpha -1}}, which
evaluates to \teq{\simeq 0.15} for a polar cap size of \teq{\Theta
=5^\circ} and \teq{\alpha =1.6}.  Reducing \teq{\erg_{\max}} clearly
diminishes the proportion of photons above pair threshold.

In the presence of photon splitting, above \teq{B_0 \sim 0.5 B_{\rm
cr}}, the cascade is limited to only one pair generation
in each polarization mode, amounting to one pair per \teq{\parallel}
mode photon injected above the pair escape energy plus two pairs per
\teq{\perp} mode photon injected above the splitting escape energy (see
Figure~\ref{fig:paircas}).  The results clearly indicate that photon
splitting in only one mode does not significantly suppress the pair
yield.  As long as one photon polarization mode does not undergo
splitting, there is always a channel for pair production.  In fact,
photon splitting in one mode can even increase the pair yield, as is
evident in Figure~\ref{fig:py1mode} for field strengths between
\teq{0.2 B_{\rm cr}} and \teq{0.7 B_{\rm cr}}, for \teq{\erg_{\rm max}
\ge 10^4}.  Consequently, the putative radio quiescence boundaries in
Fig.~\ref{fig:quiescence} are not borders to domains
of significant pair suppression if photons of \teq{\parallel}
polarization cannot split.  While splittings
\teq{\perp\to\parallel\parallel} can compete effectively with pair
creation, the two produced \teq{\parallel} photons can only create
pairs if splitting of \teq{\parallel} photons is forbidden by kinematic
selection rules.  In such a case, pair creation is prolific unless the
second generation \teq{\parallel} photons are around or below the
escape energy for pair production, \teq{\erg_{\rm esc}^{\parallel\to
e^+e^-}}, i.e. the original \teq{\perp} photon is of energy
\teq{\lesssim 2\erg_{\rm esc}^{\parallel\to e^+e^-}}.  Hence, in the
energy range \teq{\erg_{\rm esc}^{\parallel\to e^+e^-}\lesssim
\erg\lesssim 2\erg_{\rm esc}^{\parallel\to e^+e^-}},
\teq{\perp\to\parallel\parallel} splittings truly do suppress pair
creation; otherwise they merely postpone it a generation.  By comparing
the number of primary photons in this range with that above
\teq{\erg_{\rm esc}^{\parallel\to e^+e^-}}, it is quickly estimated
that the pair yield is enhanced by a factor of the order of
\teq{2^{2-\alpha}} when splitting is switched on, i.e.  generally a
minimal change.  This is precisely the behaviour observed in
Figure~\ref{fig:py1mode} for all but the \teq{\erg_{\rm max}=10^2}
case.  Effectively, the doubling of the number of photons by splittings
compensates almost exactly for the suppression of pair creation below
\teq{2\erg_{\rm esc}^{\parallel\to e^+e^-}}.  Therefore, when only
\teq{\perp} mode photons can split, pair creation is inhibited only
when \teq{\erg_{\rm max}\lesssim 2\erg_{\rm esc}^{\parallel\to
e^+e^-}}.

The situation changes dramatically if three modes of photon splitting
(those allowed by QED) are operating.  As shown in
Figure~\ref{fig:py1mode} for the case of \teq{\erg_{\rm max} = 10^2}
and \teq{\erg_{\rm max} = 10^4} (the other cases not shown behave
similarly to this case), the pair yield drops dramatically above \teq{B
\sim 0.4 B_{\rm cr}} (depending on \teq{\Theta}) when photons of both
polarization modes are allowed to split.  Pure photon splitting
cascades can now take place, and typically exceed ten photon
generations.  In fields above \teq{B_{\rm cr}} very few pairs are
created because the photon splitting cascade degrades the photon
energies below the pair escape energy before the cascade reaches a
height where pair production can take over.  Figure~\ref{fig:py3modes}
shows the effect of varying the primary photon injection colatitude on
the cascade pair yield, as a function of magnetic field strength for
cascades where three modes of splitting are allowed.  The field
strength at which the pair yield begins to drop increases with
decreasing colatitude, due to the increasing field line radius of
curvature.  This effect was seen in the radio quiescence boundary,
where \teq{B_0 \propto \Theta^{-4/15}} was computed in
Section~\ref{sec:quiescence} by equating pair and splitting escape
energies.  A similar boundary may be predicted by equating the
polarization-averaged escape energies for pair creation and photon
splitting, with the result being a virtually identical dependence of
\teq{B_0} on \teq{\Theta}, but with a constant of proportionality that
is a factor of about 1.7 lower than that in Eq.~[\ref{eq:deathline}].
Such a polarization-averaged boundary of radio quiescence reproduces
the (\teq{B_0}, \teq{\Theta}) phase space corresponding to the
precipitous declines observed in Figure~\ref{fig:py3modes}.  Hence it
can be concluded that {\it the simplified criterion for pair
suppression by splitting is fairly accurate if all three QED-permitted
modes of splitting are operating}.

\figureout{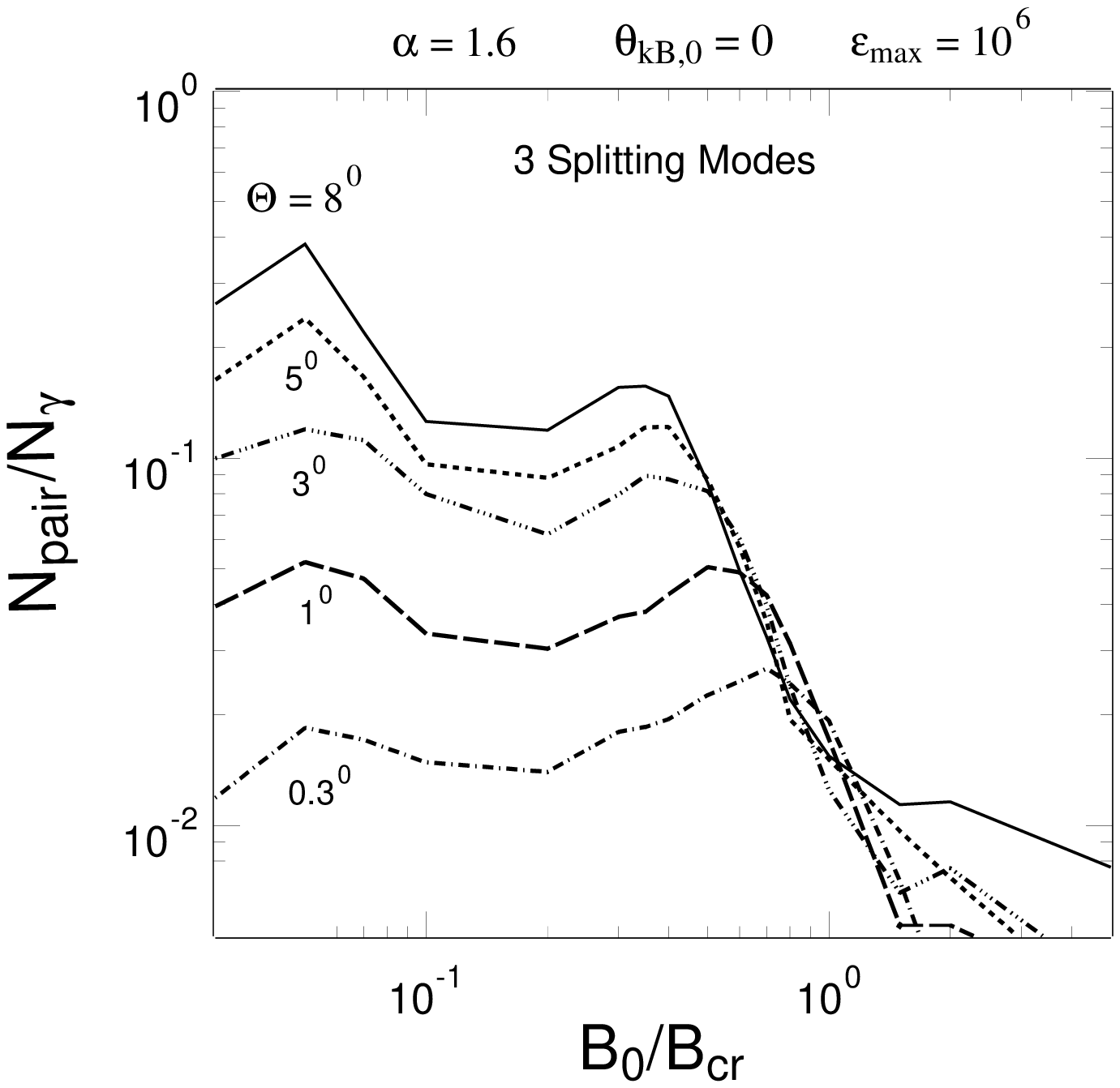}{4.9}{1.0}{-0.2}{
 The cascade pair yield (number of pairs per injected photon) as a
 function of surface magnetic field strength, \teq{B_0}, in units of the
 critical field, \teq{B_{\rm cr}}, for different primary photon
 injection colatitudes, \teq{\Theta}, for the case where three photon
 splitting modes operate.  Pair suppression becomes substantial in
 near-critical and supercritical fields when pure photon splitting
 cascades operate.  Fluctuation in the curves at the highest field
 strengths is numerical noise due to low counts. Again, surface
 emission of photons is assumed.
\label{fig:py3modes}}            

\begin{figure*}[t]
\centerline{}
\centerline{\hskip 0.45truein\psfig{figure=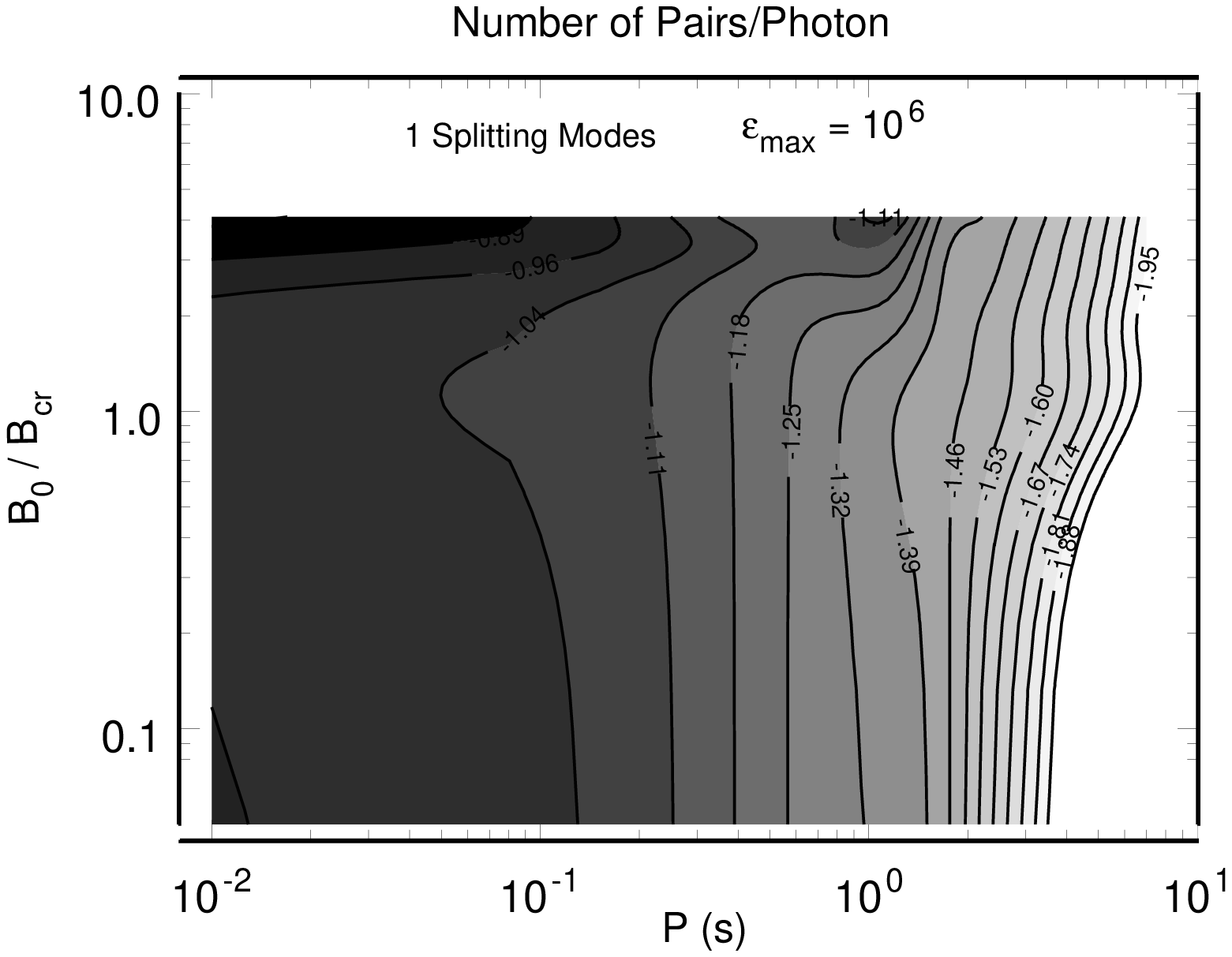,width=4.5in}
        \hskip -0.5truein \psfig{figure=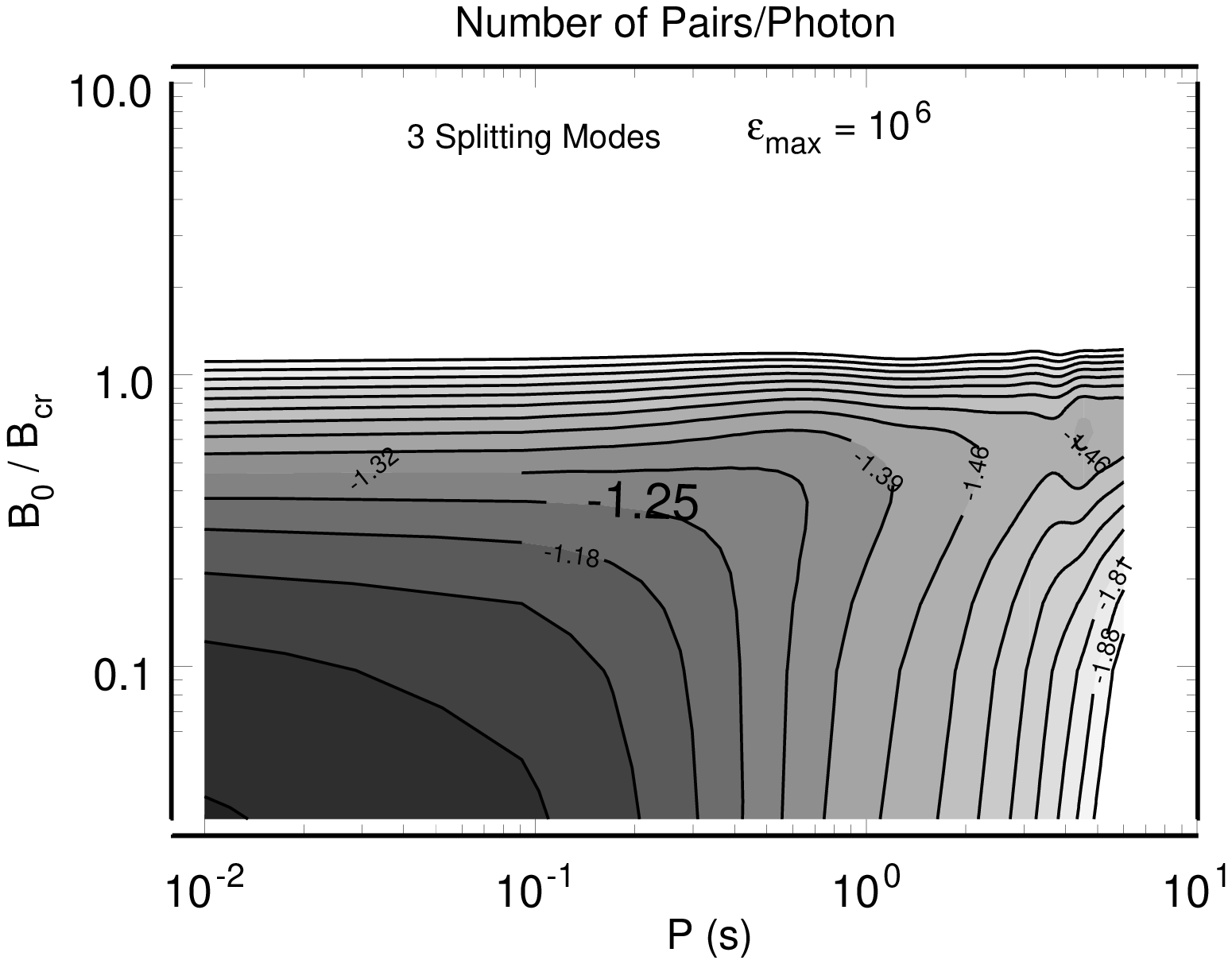,width=4.5in}}
 \vspace{-0.3in}
 \figcaption{
 Contour plots of cascade pair yield (number of pairs per injected
 photon) as a function of surface magnetic field strength, \teq{B_0}, in
 units of the critical field, \teq{B_{\rm cr}}, and pulsar period for
 the cases where (left panel) only one mode of photon splitting is
 allowed, and (right panel) where three photon splitting modes are
 permitted.  Contour levels are equally spaced logarithmic (base ten)
 intervals.
 \label{fig:contour}}          
\end{figure*}

As a more complete representation of the relevant phase space, contour
plots of the pair yield as a function of magnetic field and pulsar
period are shown in Figure~\ref{fig:contour} for the cases of one and
three splitting modes.  The period is derived from the surface
injection colatitude assuming emission at the rim of a polar cap of
standard size \teq{\Theta} as given in Eq.~(\ref{eq:polarcap}).  In the
left panel of Figure~\ref{fig:contour}, which shows the case for one
splitting mode, the most dramatic feature is the sharp fall-off of pair
yield at large pulsar periods.  This effect is purely a consequence of
the polar magnetic field geometry: at large periods (small injection
colatitudes) the radius of curvature of the field lines is larger and
the pair production attenuation length can exceed a stellar radius,
suppressing pair cascades.  This defines the ``death line" for radio
pulsars. There is a weak dependence of the ``death line" on magnetic
field strength for \teq{B \gtrsim 0.1 B_{\rm cr}} because the pair
attenuation length saturates due to threshold pair production. 
With only one splitting mode in operation, the
pair yield has a relatively weak variation with magnetic field.
However, there is a maximum in pair yield at fields \teq{B
\lesssim 0.1 B_{\rm cr}} and short periods, because pairs are produced
in high Landau states.

In the right panel of Figure~\ref{fig:contour}, which shows the case
where three splitting modes are operating, the most dramatic feature is
a sharp fall-off of pair yield at magnetic fields around \teq{B \sim
B_{\rm cr}}.  This feature is the radio quiescence line due to photon
splitting as described in Section~\ref{sec:quiescence}  The period
dependence of this line is consistent with that inferred
from the escape energy calculation, i.e.  \teq{B_0 \propto
\Theta^{-4/15} \propto P^{-2/15}}.  The radio ``death line" is also
apparent as a decline in pair yield at large periods.

\section{DISCUSSION}
 \label{sec:discussion}

\subsection{Radio quiescence?}
 \label{sec:radio}

The results of the previous section show that if three modes of
splitting operate at high field strengths, then there is nearly
complete pair suppression at high field strengths and the approximate
condition for radio quiescence proposed in Section~\ref{sec:quiescence}
is valid. If only one mode of splitting (i.e.
\teq{\perp\to\parallel \parallel}) operates, then some other effect
must act to reduce the pair yield, otherwise, pair creation in cascades
remains prolific at near-critical and supercritical fields, and there
is no reason to expect a regime of radio quiescence in the upper
portion of the \teq{P}--\teq{\dot P} diagram.  An answer to the
question of how many modes of photon splitting operate in highly
dispersive magnetic fields is not presently available and will require
detailed investigation into the behavior of the photon splitting rate
in this regime.

We have discussed several ways in which a suppression of pair creation 
at high B can be present if only one splitting mode operates.  First, 
if the effective maximum energy
\teq{\erg_{\max}} of primary photons satisfies \teq{\erg_{\max}\lesssim
2\erg_{\rm esc}^{\perp\to e^+e^-}}, then the pair yield must be
dramatically reduced whenever such photons are mostly of \teq{\perp}
polarization.  However, there is no reason {\it a priori} to believe that such
low values of \teq{\erg_{\max}} should arise when \teq{B_0\gtrsim
B_{\rm cr}}.  For curvature radiation, the spectrum is only dependent
on \teq{B_0} via acceleration properties.  For resonant Compton
scattering, the spectral dependence on \teq{B_0} is more complicated
and needs to be the subject of future investigation.  While the maximum
electron energy declines somewhat with \teq{B_0} (e.g.  Harding \&
Muslimov 1998; see the discussion just below), the primary photon
spectrum is flat up to \teq{\sim B^2/\erg_s} (for
soft target photon energy \teq{\erg_s} and drops steeply thereafter:
Dermer 1990; Baring 1994).  Hence, properties of the primary spectrum
are unlikely to uniformly produce suppression of pair creation at
high field strengths.

A second possibility that is effective at diminishing, but not
completely suppressing, the number of pairs when \teq{\erg_{\max}\gg
\erg_{\rm esc}^{\perp\to e^+e^-}} is ground state pair creation.  It
inhibits successive generations of pair cascading and becomes
significant in local fields \teq{B\gtrsim 2\times 10^{12}}Gauss.  This
value is below the spin-down fields of the most magnetic radio pulsars,
so that evidently ground state pair creation cannot be primarily
responsible for imposing radio quiescence.  Yet this physical effect
could be a contributory factor, noting that the extant pulsar
population is somewhat thinner in the \teq{P}--\teq{\dot P} diagram
than might be expected at \teq{B_0\gtrsim 10^{13}}Gauss.  A related
possibility (as discussed in Section~\ref{sec:suppposit}) is that
positronium formation may inhibit radio emission without reductions in
pair creation.  But even if stable positronium formation does occur in
pulsar magnetospheres, it cannot account for observed radio quiescence
in pulsars because the predicted onset of positronium formation occurs
at field strengths \teq{\sim 0.1 B_{\rm cr}}, in the middle of the
radio pulsar population.  Indeed, Usov \& Melrose (1996) invoke
bound-state pair creation in their model for gamma-ray emitting radio
pulsars.

Finally, we have argued that suppression of pair creation by the
conversion of \teq{\parallel} polarization photons that are the product
of \teq{\perp\to\parallel\parallel} splittings back into the
\teq{\perp} state by the resonant Compton scattering process, before
they can pair produce, will not be effective in normal pulsar
magnetospheres where the particle densities are low.

While these theoretical issues leave the question of radio quiescence
quite open, several observational issues are pertinent to this
discussion.  These relate to whether or not one expects to observe
radio pulsars with high \teq{\dot P}, which divides into issues of
intrinsic population densities in the \teq{P}--\teq{\dot P} diagram and
observational selection effects.  There are several obvious natural
biases influencing pulsar observability.  First, due to the expectation
that pulsars with higher spin-down power (i.e. \teq{B_0^2/P^4}) are
probably more luminous, a property observed in both the X-ray (Becker
\& Tr\"umper, 1997) and gamma-ray (e.g. Zhang \& Harding 2000) bands,
one might anticipate that high-field radio pulsars are more luminous
than their less-magnetized counterparts.  There is evidence that this
trend is present in the observed population: when luminosity
distributions for Princeton Pulsar Catalog members are binned in ranges
of \teq{B_0}, radio luminosities \teq{L_{\rm rad}} increase with
spin-down field up to around \teq{2\times 10^{13}}Gauss, beyond which
the trend appears to reverse and \teq{L_{\rm rad}} perhaps begins to
decline with \teq{B_0}.  While this reversal is suggestive of an onset
of radio-quiescence at high fields, it is presently statistically
insignificant.  The addition of the Parkes Multi-Beam population (for
which distances have not yet been established from the dispersion
measures) will improve statistics, but perhaps still leave the evidence
for a luminosity decline inconclusive.  Uncertainty in source distances
pervades estimates of pulsar luminosity, and since there is a
correlation between distance out of the galactic plane and pulsar age
(and therefore period), we caution against over-interpretation of
\teq{L_{\rm rad}} trends for high \teq{B_0}.

The next factor is the age distribution in the \teq{P}--\teq{\dot P}
diagram.  Clearly, evolution (with or without field decay) guarantees a
denser population at long periods, and the speed at which the period
evolves is a rapidly increasing function of \teq{B_0}.  The population
density would be constant along diagonal lines of constant age
\teq{P/(2{\dot P})} for a uniform birth rates vs. field distribution.
Unless there are many fewer pulsars born with high field strengths, the
smaller average age of high field pulsars to the left of the
conventional death lines cannot account for the dearth of radio pulsars
at these fields.  Concomitantly, the expected clustering of pulsars
near the death line at subcritical fields is not realized; the beaming
of radio emission is obviously a contributor to this effect.  The
reduction of the polar cap size with period should reduce the solid
angle (\teq{\propto P^{-1}}) of the cone of emission to more or less
compensate the age clustering effect.  Such a property was invoked by
Young, Manchester \& Johnston (1999) to argue that the 8.5 second
Parkes pulsar PSR J2144-3933 was not isolated in its existence, but
rather representative of a large, unseen population that challenges
conventional theories explaining the death line.  Yet the observed {\it
thinning} of the population near this boundary argues that another
factor is influential in determining the period distribution; a
reduction in luminosity due to the onset of pair quenching may provide
a partial explanation of this dilution.  Notwithstanding, the beaming
phenomenon impacts the \teq{P} distribution and not \teq{\dot P} phase
space.

There are two main observational selection effects, the first being an
intrinsically greater sensitivity at longer periods and lower
dispersion measures, basically due to interstellar dispersion and
scintillation effects, radiometer noise, and pulse shape and Fourier
analysis properties (e.g. see Cordes and Chernoff 1997 for a discussion
of sensitivities).  The second true selection effect, less evident in
the literature, is that pulsar surveys tend to filter out baseline
fluctuations on long (i.e. \teq{\gtrsim 5} second) timescales, thereby
selecting against supersecond period pulsars (Cordes, private
communication).  This bias is partly driven by past needs to focus on
the appropriate range of periods of conventional pulsars, and will be
counterbalanced by the increasing scientific interest in magnetars,
discussed in Section~\ref{sec:magnetars} below.  In summation, in the
light of all these observational considerations, without performing an
extensive population statistics analysis, it is fairly safe to argue
that radio pulsars with \teq{\dot P} in excess of \teq{10^{-11}} are
either rare or non-existent.  The confirmation of a possible radio
pulsar counterpart to SGR 1900+14 (discussed in
Section~\ref{sec:magnetars}), or otherwise, shall play a prominent role
in resolving this issue.

\subsection{Particle Acceleration Locales}
\label{sec:accel}

Since the strength of the local magnetic field is one of the key
influences on the location of the high-field death line, the site of
the high energy emission in the pulsar magnetosphere is critically
important.  The initial angle at which the high-energy photons are
emitted relative to the local magnetic field has also been shown in the
previous subsection to be of critical importance (see
Figure~\ref{fig:quiescence}).  Both of these factors are determined by
the nature of the particle acceleration above the polar cap.  Several
types of models have studied pulsar acceleration due to charge deficits
at different locations in the magnetosphere.  Polar cap models consider
the formation of a parallel electric field in the open field region
near the magnetic poles, while outer gap models consider acceleration
in the outer magnetosphere, near the null charge surface (see Mestel
1998 for the most recent and comprehensive review of pulsar
electrodynamics).

Polar cap models for pulsar high-energy emission are all based on the
idea, dating from the earliest pulsar models of Sturrock (1971) and
Ruderman \& Sutherland (1975; hereafter RS75), of particle acceleration
and radiation near the neutron star surface at the magnetic poles.
Within this broad class, there is a large variation, with the primary
division being whether or not there is free emission of particles from
the neutron star surface.  This question hinges on whether the surface
temperature \teq{T} of the neutron star (many of which have now been
measured in the range \teq{T \sim 10^5 - 10^6} K; Becker \& Tr\"umper
1997) exceeds the ion, \teq{T_{\rm i}} and electron, \teq{T_{\rm e}},
thermal emission temperatures.  If \teq{T < T_{\rm i}}, a vacuum gap
will develop at the surface, due to the trapping of ions in the neutron
star crust (RS75, Usov \& Melrose 1995).  In this case, the particle
acceleration and radiation will take place very near the neutron star
surface.  If \teq{T > T_{\rm e}}, free emission of particles of either
sign of charge will occur.  The flow of particles is then limited only
by space charge (SCLF case), and an accelerating potential will develop
(Arons \& Scharlemann 1979; Muslimov \& Tsygan 1992) due to an
inability of the particle flow all along each open field line to supply
the corotation charge (required to short out \teq{E_{\parallel}}).  In
space charge-limited flow models, the accelerating \teq{E_{\parallel}}
is screened at a height where the particles radiate \teq{\gamma}-rays
that produce pairs.  This so-called {\it pair formation front} (PFF)
(e.g.  Arons 1983, Harding \& Muslimov 1998) can occur at high
altitudes above the polar cap.

Zhang \& Harding (2000b) propose that, if both photon polarization
modes can undergo splitting at high fields, the radio quiescence line
will be much lower for pulsars with vacuum gaps (``anti-pulsars'' in
the language of RS75), than for pulsars with SCLF gaps (``pulsars'').
In the case of vacuum gaps the high energy radiation occurs near the
neutron star surface, where the local fields are high, whereas in the
case of SCLF gaps, the high energy emission (and subsequent pair
creation) may occur at high altitudes, especially when photon splitting
prevents pair creation and thus a PFF near the surface.  The radio
quiescence line for ``anti-pulsars'' will be the photon splitting death
line for surface emission, while the radio quiescence line for
``pulsars'' will be located around $B_0 = 2\times 10^{14}$ G (for a
surface temperature of $10^6 K$), above which thermionic emission of
electrons from the surface is no longer possible and ``pulsars'' must
have vacuum gaps.  This could account for the observed existence of
radio-loud pulsars such as PSR J1814 and radio-quiet pulsars such as
AXP 1E 2259 at the same field strength and in close proximity in the
\teq{P}--\teq{\dot P} diagram.

\subsection{Magnetars}
 \label{sec:magnetars}

The importance of a boundary for radio quiescence to the study of the
radio pulsar population is obvious.  Yet the absence of radio pulsars
in the high-\teq{B_0} region of phase space has become of even greater
importance with the mounting observational evidence at X-ray and
$\gamma$-ray energies for the existence of neutron stars with such
large fields, which {\it a priori} are not discriminated against (for a
given age) by known radio selection effects (as discussed in
Section~\ref{sec:radio}).  This evidence includes the detection of
spin-down in the growing number of {\it anomalous X-ray pulsars}
(AXPs).  These sources have been known to exist for over a decade (e.g.
see the biographical summary of Mereghetti \& Stella 1995), though the
identification of them as being anomalous was forged slowly with the
accumulation of observational data.  The designation ``anomalous,''
coined by van Paradijs, Taam \& van den Heuvel (1995), was founded in
their periods (6--12 seconds) being relatively short compared with
typical accreting X-ray binary systems, combined with their unusually
steep X-ray spectra and monotonic increases in periods:  short period
binary X-ray pulsars usually exhibit \teq{\dot{P}<0}, i.e. spin-up.  A
list of AXPs with measured \teq{P} {\it and} \teq{\dot{P}} is given in
Table~1, an adaptation of that in Gotthelf \& Vasisht (1998).
Furthermore, the association of two of these pulsars (1E 1841-045 and
1E 2259+586) with supernova remnants (which are typically much younger
than X-ray binaries) has shifted the focus from accretion torques, as
championed by Mereghetti \& Stella (1995) and van Paradijs, Taam \& van
den Heuvel (1995), to electromagnetic dipole torques (e.g. Vasisht \&
Gotthelf 1997) as the origin of the spin-down.  In fact, Vasisht \&
Gotthelf (1997) argue that it is difficult for accretion torques to
spin a pulsar down to \teq{P\sim 10} seconds on a \teq{10^3}year
timescale unless the pulsar was born a slow rotator.

\begin{figure*}[t]
\centerline{}
\centerline{\psfig{figure=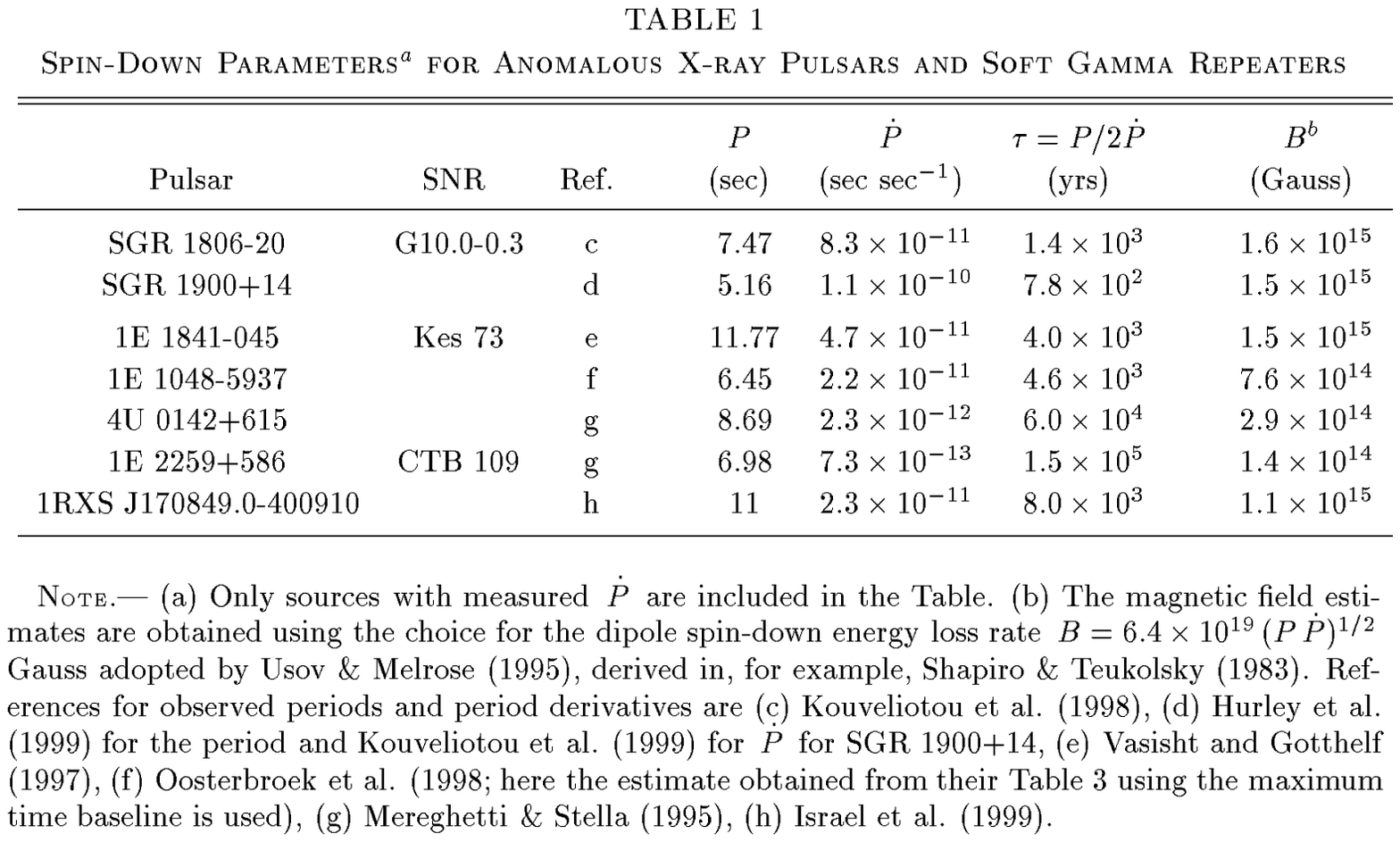,width=6.2in}}
\end{figure*}

The electromagnetic dipole interpretation for the spin-down immediately
implies immense supercritical (\teq{B_0> 4.41\times 10^{13}}Gauss)
fields in these sources: inferred values are given in Table~1.  The
motivation for such a perspective has been dramatically enhanced by the
recent detection of similar periods and period derivatives of
soft gamma repeaters (SGRs), the neutron star sources with transient
outbursts of soft gamma-ray emission possessing quasi-thermal spectra.
Duncan \& Thompson (1992) postulated that neutron stars with
supercritical magnetizations, {\it magnetars}, were responsible for SGR
activity based on the observed 8 second periodicity (Mazets et al.
1981) of the 5th March 1979 outburst from SGR 0525-66, combined with
it's strong directional association (Helfand \& Long 1979; Cline et
al.  1982) with the supernova remnant N49 in the Large Magellanic
Cloud.  No spin-down \teq{\dot{P}} was measurable for this source since
the data that unambiguously exhibited periodicity spanned a single
range of around two minutes.  Hence the surface field of \teq{B_0\sim
6\times 10^{14}}Gauss inferred by Duncan \& Thompson (1992) for SGR
0525-66 was purely circumstantial.  Nevertheless, their hypothesis was
dramatically bolstered by the detection of a supersecond periodicity
{\it coupled with a high rate of spin-down} in the quiescent
counterparts of SGR 1806-20 (Kouveliotou, et al.  1998) and of
SGR1900+14 (Kouveliotou et al.  et al. 1999); parameters for these
sources are also listed in Table~1.  A giant burst, very similar to
that observed from SGR 0525-66, was seen from SGR1900+14, exhibiting
the same 5 s period as was detected in the quiescent emission (Hurley
et al. 1999), strengthening the magnetar identification for SGRs.
While the similarity of \teq{P} and \teq{\dot{P}} and the existence of
remnant identifications for both AXPs and SGRs may provide moderately
compelling motivation for classifying the two types of sources
together, given the vastly different emission properties between AXPs
and SGRs, their commonality may be confined just to magnetars being the
sites for their activity.

\begin{figure*}[t]
\centerline{}
\centerline{\psfig{figure=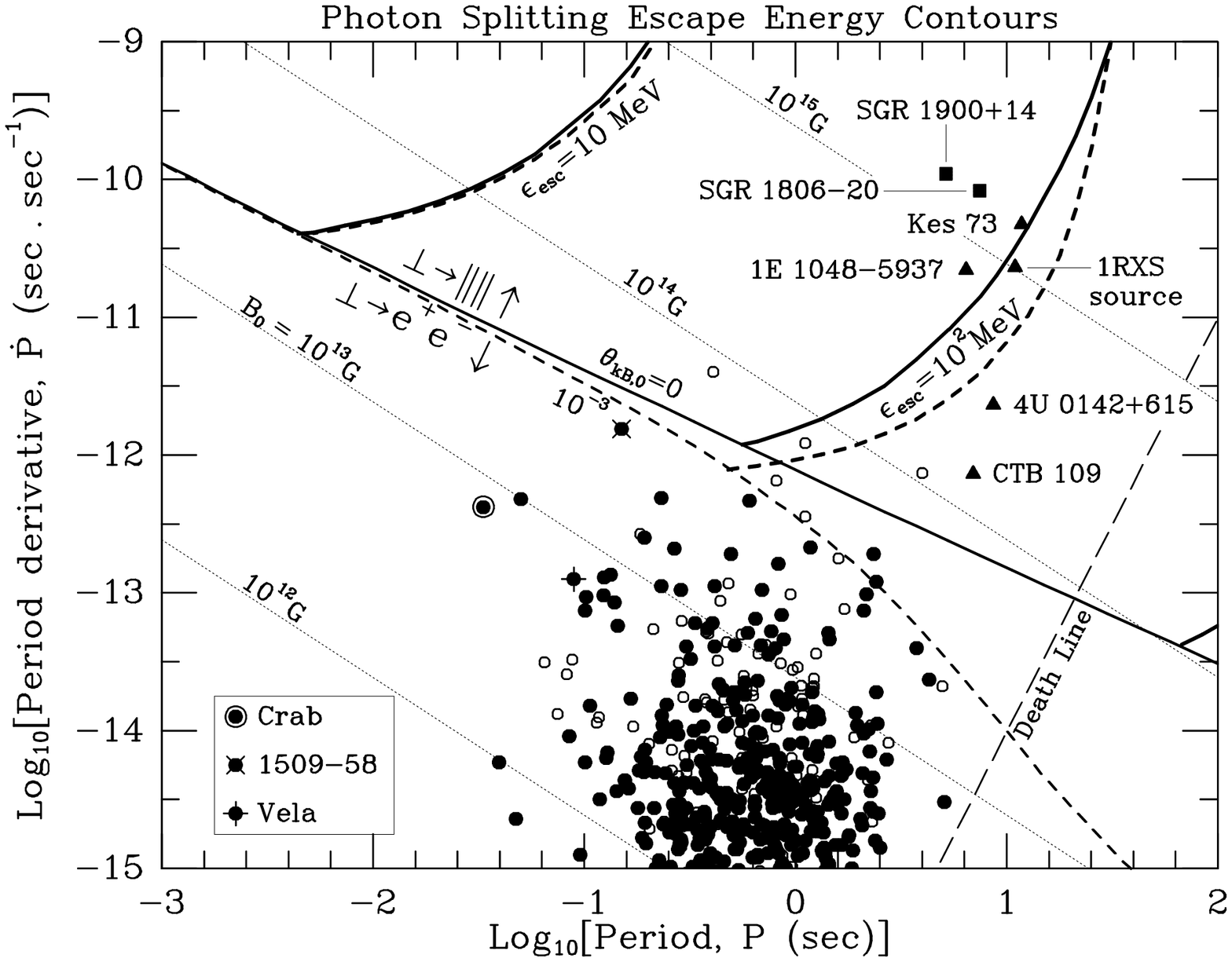,width=5.9in}}
\figcaption{
The upper region of the \teq{P}-\teq{\dot{P}} diagram, with filled
circles again denoting the locations of members of the latest edition
of the Princeton Pulsar Catalogue, and the open circles marking pulsars
in the recent Parkes Multi-Beam survey.  The Crab, Vela and PSR 1509-58
gamma-ray pulsars are highlighted as indicated in the lower left inset,
together with the positions of five radio-quiet anomalous X-ray pulsars
(open square stars) and SGRs 1806-20 and 1900+14 (filled squares) in
the upper right of the diagram; their measured \teq{\dot P} and
inferred fields \teq{B_0} are listed in Table~1.  The conventional
death line (as in Fig.~\ref{fig:quiescence}) lies at periods longer
than all seven of the putative magnetars.  The \teq{\theta_{\rm
kB,0}=0} (midweight solid diagonal curve) and \teq{\theta_{\rm
kB,0}=10^{-3}} (midweight dashed curve) depictions of the radio
quiescence boundary from Fig.~\ref{fig:quiescence} are exhibited.  The
heavy-weight pairs of curves labelled \teq{\erg_{\rm max}=10}MeV and
\teq{\erg_{\rm max}=10^2}MeV are contours for the escape energy of
photon splitting, \teq{\perp\to\parallel\parallel}, with the solid and
dashed line styles denoting \teq{\theta_{\rm kB,0}=0} and
\teq{\theta_{\rm kB,0}=10^{-3}}, respectively.  They are shown only
above the potential radio quiescence boundaries, where splitting
dominates pair creation for the \teq{\perp} polarization, marking
contours for the approximate maximum observable energy permitted in
highly-magnetized pulsars; to the right of these the magnetosphere is
transparent to 10 MeV and 100 MeV photons, respectively.
 \label{fig:escape_cont} }      
\end{figure*}

A property mutual to anomalous X-ray pulsars and soft gamma repeaters
is that they are radio quiet pulsars (if one excepts the unconfirmed
detection of SGR 1900+14 by Shitov 1999), an intriguing fact given
their high magnetic fields and the focus of this paper.  The spin-down
parameters for these sources (see Table 1), are used to derive their
positions on the \teq{P}--\teq{\dot P} diagram in
Figure~\ref{fig:escape_cont}.  The AXPs and SGRs all lie in the extreme
upper right of the \teq{P}--\teq{\dot P} phase space, above our derived
$\theta_{\rm kB}$ radio quiescence line.  However, one of the radio
pulsars recently discovered in the Parkes Multi-beam Survey, PSR
J1814-1744, lies very close to the AXP CTB 109, such that a single
radio quiescence line cannot separate the radio pulsar and AXP
populations.  This proximity coupled with the fact that PSR J1814-1744
has not been detected in X-rays (Pivovaroff, Kaspi \& Camilo 2000)
strongly suggests that a quantity other than \teq{P} and \teq{B_0} has
a profound influence on the properties of highly-magnetized radio
pulsars and AXPs.  Note that there are several effects which could
cause a fuzziness in both the radio quiescence boundary and the
positions of the pulsars in the \teq{P}--\teq{\dot P} diagram.  As
mentioned in Section \ref{sec:quiescence}, the effective quiescence
boundary for a particular source depends on the altitude of emission.
The quiescence boundary applicable for high-B radio pulsars, whose
high-energy emission may occur at some altitude above the surface (see
Section \ref{sec:accel}), may be higher than the quiescence boundary
applicable for AXPs, whose emission is thought to be predominantly of
thermal origin from the hot neutron star surface.  Another possibility,
pertinent if the radio detection of SGR 1900+14 is confirmed, is that
SGRs are ``pulsars'' with active accelerators allowing pair production
and radio emission at high altitudes, as discussed by Zhang \& Harding
(2000b, see Section \ref{sec:accel}), while AXPs are radio-quiet
``anti-pulsars''.

Furthermore, the actual magnetic fields of the AXP and SGR sources are
not likely to be those derived from pure dipole spin-down.  If rapid
field decay is powering the luminous emission from these sources
(Thompson \& Duncan 1996), then it is likely that transient higher
multipoles exist near the surface.  Since higher multipole radiation
contributes relatively less to the spin-down torque, multipole fields
significantly stronger than those derived assuming a pure dipole could
be present at the surface without affecting the spin-down.  This would
effectively lower the radio quiescence boundary for the AXPs, both due
to the increase in field strength and decrease in radius of curvature.
Measurement of braking indices larger than 3 (the dipole value) in
these sources may reveal the presence of higher multipoles.  In fact, a
marginally significant detection of $\ddot\nu$ in the AXP 1RXS
J170849.0-400910 (Kaspi, Chakrabarty \& Steinberger 1999) would give a
braking index in excess of \teq{100}!  Field decay alone would not
change estimates of the surface field, but would decrease the age of
the pulsar relative to the dipole value (Colpi, Geppart \& Page 1999).

On the other hand, the spin-down of some magnetars could be influenced
by powerful particle winds, in this case lowering the estimated surface
field strength and increasing the characteristic age (Harding,
Contopoulos \& Kazanas 1999, Thompson et al. 1999) because the wind
increases the spin-down rate by distorting the dipole field inside the
light cylinder.  This is more likely to be a factor in SGRs, where
there is evidence for particle emission associated with bursts (Frail,
Vasisht \& Kulkarni 1997; Frail, Kulkarni \& Bloom 1999).  Wind
luminosities in excess of \teq{\sim 10^{36}\,\rm erg \,s^{-1}} in
SGR1806-20 would lower the derived surface field strength below
$10^{14}$ G.  An episodic wind, which is more likely, would decrease
the estimated surface field to a lesser degree.  Possibly for this
reason, the dipole surface fields of the SGRs are higher than those of
most of the AXPs and more than an order of magnitude higher than the
most strongly magnetized radio pulsars.

\subsection{Guides for Soft and Hard Gamma-Ray Pulsation Searches}
 \label{sec:guide}

To close this section of discussion, it is appropriate to mention the
implications of our analysis for gamma-ray pulsar searches, which are,
in a sense, independent of the radio quiescence issue.  If the
electrons are accelerated to high Lorentz factors in profusion, then
primary photons will extend into the EGRET band and beyond.  Therefore,
short period pulsars with high spin-down power (\teq{\propto
B^2_0/P^4}) like the Crab pulsar should be luminous and therefore
easily detectable by future experiments such as GLAST.  At longer
periods near the magnetar regime, the collected observational data and
the age distribution bias (see Section~\ref{sec:radio}) indicate that
spin-powered pulsars will be generally faint unless they are nearby.
Yet the observed hard X-ray luminosities of AXPs are well in excess of
their spin-down luminosities so that non-rotational energy sources are
indicated.  Hence luminosity biases against long period pulsars may not
be as pronounced for those with alternative power sources such as
magnetic field decay, at least in the magnetar regime.
The work of this paper clearly underlines the fact
that photon attenuation considerations have as profound an influence on
gamma-ray observability as do issues of power in the primary
radiation.   

At near-critical and supercritical fields, the attenuation by pair
creation and photon splitting in the EGRET band is dramatic, as is
evident from the contours of maximum energy depicted in
Figure~\ref{fig:escape_cont}.  Basically, photon transparency at a
given energy is achieved to the right of and below (i.e. at longer
periods and lower \teq{\dot P}) a given contour so that the positioning
of the \teq{\erg_{\rm esc}=100}MeV splitting escape energy contours
indicates that it will not be easy for an experiment like GLAST
(depending on its final design) to detect many or most of the known
magnetars, and the new Parkes Multi-Beam sources PSR J1119-6127 and PSR
J1726-3530.  Should magnetars emit in the gamma-rays, they can only do
so generally below the EGRET band.  This conclusion holds even if the
magnetospheric field structure is non-dipolar, since escape energies
are pushed to lower values by greater field curvature.  A possible
evasion of this constraint is that the inferred fields are significant
overestimates of the true surface fields, as might occur for wind-aided
spin-down in magnetars (see Section \ref{sec:magnetars}), so that the
escape energy contours may possess pathological distortions in the
\teq{P}-\teq{\dot P} diagram.  Moreover, such a two-dimensional phase
space may be insufficient to account for the magnetospheric and photon
attenuation properties.  Hence high energy astrophysicists interested
in pulsar searches at high \teq{\dot P} should extend their period
range to include supersecond periods to maximize their potential
harvest.  While this range has historically been given low priority
(e.g. see the search in Mattox et al. 1996), the excitement generated
by AXP and SGRs in recent years is changing emphases in such search
programs.

These attenuation properties emphasize that spectroscopy in a variety
of gamma-ray pulsars will probably provide the ability to discriminate
between the applicability of polar cap or outer gap models.  In the
polar cap picture, the analysis of this paper shows that spectral
cutoff energies possess a strong inverse correlation with surface field
strength \teq{B_0} and the polar cap size \teq{\Theta}.  For Crab-like
and Vela-like pulsars such cutoffs are coupled to magnetic pair
creation, and appear in the EGRET band.  However, for highly-magnetized
pulsars such as PSR 1509-58, such properties are ideally explored with
a medium energy gamma-ray experiment, and probe the action of photon
splitting.  Expectations for trends of gamma-ray cutoffs in outer gap
models are not as well studied.  Yet, it appears that their trends with
\teq{B_0} and pulsar period should differ significantly from those for
polar cap models, due to the inherently different physics involved.
Cutoffs in the outer gap scenario represent maximum energies of
acceleration rather than the effects of photon absorption, and so may
actually be independent of or increase with surface field strength, and
decline with pulse period, contrary to the indications of the polar gap
model.  Such distinctive predictions can be probed by significant
population datasets such as those to be afforded by the GLAST mission.
A related observational diagnostic is spawned by the contention (Chen
\& Ruderman 1993) that gamma-ray emission is not expected in outer gap
models when the pulsar period exceeds a sizeable fraction of a second.
Hence detection of gamma-ray pulsations from a source with a
supersecond period would clearly favor the polar cap model.

Our studies also indicate that the gamma-ray spectra of pulsars should
produce distinctive polarization signatures for the polar cap model.
This should enable discrimination from outer gap scenarios in a future
era when gamma-ray polarimetry is possible.  While the predictions of
gamma-ray polarizations might be similar for the continuum spectra in
each of these two competing models, principally because their continuum
emission processes are similar, the pair creation and photon splitting
mechanisms generate spectral cutoff energies that are
polarization-dependent (as emphasized by HBG97), and therefore
immediately distinguishable from outer gap model turnovers that have no
intimate connection to QED processes in strong fields.  In particular,
when \teq{B_0\gtrsim B_{\rm cr}}, the dominance of splittings
\teq{\perp\to\parallel\parallel} will guarantee that the cutoff energy
for \teq{\parallel} photons exceeds that for ones of \teq{\perp}
polarization.  In contrast, at lower fields strengths, photon splitting
is removed from the picture and the escape energy for pair creation by
\teq{\parallel} photons is slightly lower than that for \teq{\perp}
ones, resulting in a dominance of \teq{\perp} photons near the maximum
observable energy.  Because the strong dependence of the spectral
high-energy cutoff in fields \teq{B_0\gtrsim B_{\rm cr}} occurs
exclusively when only one mode of splitting is operative, it may be
possible to resolve by observation the question of whether one or
several modes of splitting operate in high fields.  The splitting and
pair production cutoffs are expected around 100 MeV in long period
pulsars (slightly higher if the emission region is above the surface).
If only the \teq{\perp\to\parallel\parallel} mode operates, the
splitting cutoff will occur at an energy around a factor of 2-3 lower
than the pair production cutoff.  GLAST (depending on sensitivity below
100 MeV) or a future medium energy $\gamma$-ray detector would then see
100\% $\parallel$ polarization of the highest energy photons, i.e.
those between \teq{\erg_{\rm esc}^{\perp\to\parallel\parallel}}
and \teq{\erg_{\rm esc}^{\parallel\to e^+e^-}}. On the
other hand, the absence of this signature would be consistent with, but
not necessarily imply, at least two active splitting modes.  These
distinctive signatures establish a strong case for performing gamma-ray
polarimetry experiments, an objective that may not be that far in the
future for medium-energy Compton gamma-ray telescopes.

\section{CONCLUSION}
 \label{sec:conclusion}

We have presented a detailed study of the comparative attenuation of
photons by magnetic pair production and photon splitting, of pair
cascades and of conditions for pair suppression in very highly
magnetized pulsars.  While ground-state pair creation and positronium
formation can act at \teq{B_0\gtrsim 0.1B_{\rm cr}} to reduce the
number of {\it free} pairs, only photon splitting has the potential
capability of dramatically inhibiting the creation of pairs, bound or
free.  Our quantitative study of cascade pair yields at different field
strengths confirms the location of the approximate radio quiescence
boundary in the \teq{P}-\teq{\dot P} diagram proposed by Baring \&
Harding (1998), assuming that copious pair production is a requirement
for pulsar radio emission.  However, it also shows that the existence
of such a boundary requires the rate of photon splitting to be non-zero
for both photon polarization modes.  This is not believed to be true in
the low-field, weakly dispersive limit, where kinematic selection rules
prohibit splitting of photons of \teq{\parallel} polarization.  Whether
photon splitting can significantly suppress pairs in highly magnetized
radio pulsars depends on presently unexplored physics.  It therefore
has become critically important to study the photon splitting process
in the high-field, highly dispersive regime.  Understanding the
behavior of splitting in this regime will not only resolve the radio
quiescence question, but is crucial to modeling acceleration and
radiation in high-field pulsars and magnetars.

The circumstantial evidence, that the most highly magnetized pulsars
(AXPs and SGRs) tend to be radio quiet, argues for some pair
suppression at high field strengths.  The radio pulsar and magnetar
populations are no longer cleanly separated in \teq{P}-\teq{\dot P}
space, which indicates that another dimension of phase space is
necessary to divide these two source populations.  Delineation of these
two source groups is complicated by the likelihood of alternate
spin-down mechanisms and evolution issues for magnetars.  More
systematic surveys of the overlap region of \teq{P}-\teq{\dot P} space
at X-ray and $\gamma$-ray energies, correlated with radio pulsar
surveys are needed.  We have given some guidelines for such high-energy
searches based on the expected physics of particle acceleration and
photon attenuation, in anticipation of the next generation of medium
and high-energy gamma-ray experiments such as GLAST.  Spectral and
polarization observations may be crucial in identifying signatures of
photon splitting and even in resolving the issue of which splitting
modes are operating in ultra-magnetized sources, an attractive
possibility for both astrophysicists and physicists.

\acknowledgments 
We are grateful to Peter Gonthier and Marty Knecht for help with the 
splitting cascade code development.  We also thank Alex Muslimov, Jim Cordes, 
Eric Gotthelf, David Thompson, Peter Gonthier and Matthew Bailes for 
discussions.  This work was supported by the NASA Astrophysics Theory Program.


\clearpage

\end{document}